\documentclass{aa}  
\usepackage{graphicx}
\usepackage{txfonts}

\begin{document}

   \title{Inhomogeneous magnetic coupling in exoplanets:\\ the stop \& go of WASP-18 b's atmospheric flows}

   \author{A. Blöcker
          \inst{1}
          \and
          L. Carone\inst{1}
          \and
          Ch. Helling\inst{1,2}}

   \institute{Space Research Institute, Austrian Academy of Sciences, Schmiedlstrasse 6, 8042 Graz, Austria\and
   Institute for Theoretical Physics and Computational Physics, Graz University of Technology, Petersgasse 16, 8010 Graz, Austria\\
              \email{Aljona.Bloecker@oeaw.ac.at}}

  \abstract
  {Early studies of ionization in hot Jupiter atmospheres suggest that magnetic coupling may affect their dynamics and, hence, their weather and climate states. These effects may be most pronounced in ultra-hot gas giants, assuming they generate their own global magnetic field. WASP-18~b, one of the best studied ultra-hot Jupiters, hosts a highly  ionized dayside atmosphere extending deep enough to be strongly influenced by magnetic forces.
   Phase curve observations suggest an effective magnetic drag, yet its impact on the atmospheric circulation remains poorly constrained.}
   {The aim is to explore the effect of magnetic drag in atmospheres with an inhomogeneous ionization on the local and global dynamics to ultimately provide a pathway to constrain the planet's magnetic field strength.  }
{An analytical  parameterization for anisotropic magnetic drag, including both Pedersen and Hall drag components, and associated frictional heating in the globally neutral atmosphere, is implemented in the 3D General Circulation Model \texttt{ExoRad}  to study WASP-18 b. 
  Fundamental plasma parameters are analyzed to explore where magnetic coupling becomes important in the atmosphere, depending on the  dipolar field geometry, the ionization fraction, and the collisional coupling between charged particles and neutrals.
   Climate characteristics are compared for different drag formulations, to assess whether anisotropic drag physics is required,
   for accurately capturing magnetic coupling effects.}
   {Anisotropic magnetic drag and frictional heating,  both shaped by local ionization, strongly affect  wind strength and direction  in the upper atmosphere, modifying the day-night circulation and producing observable temperature asymmetries. Anisotropic drag enhances the evening-morning terminator temperature difference at 0.1~bar, and generates two off-equator hotspots with reduced eastward shift. The terminator regions are in particular susceptible to how magnetic drag is described in the model.}
   {Anisotropic magnetic drag damps and redirects the dayside-to-nightside winds, partially decoupling the equatorial flow at the morning terminator while maintaining the  nightside jet.
   Locally changing drag forces and frictional heating create  asymmetric temperature patterns manifesting as primary and secondary hotspot regions.}


   \maketitle
\nolinenumbers
\section{Introduction}
Ultra-hot Jupiters (UHJs) are tidally locked gas giants with dayside gas temperatures exceeding 2000–3000 K, resulting in partially ionized atmospheres under intense stellar irradiation (e.g., \citealt{2019A&A...631A..79H}). 
At sufficiently high gas temperatures, alkali metals and other species thermally ionize, leading to a weakly ionized plasma in the atmosphere \citep{Lodders2003,RodriguezBarrera2015,2019A&A...631A..79H,Dietrich2022}. 
Since the atmospheric gas behaves as a plasma on the dayside, it interacts electromagnetically with the planetary magnetic field. 
The resulting magnetic drag force has been suggested as a key mechanism for damping zonal jets and contributing to localized heating in these atmospheres \citep[e.g.,][]{Perna2010, Batygin2013}.
The role of magnetic fields in shaping their atmospheric circulation remains a challenging problem and has significant implications for the (U)HJ's climate dynamics and wind patterns. 

A major uncertainty is the knowledge about the actual magnetic field strengths of (U)HJs. While scaling laws based on self-sustained dynamo simulations can predict fields of a few to hundreds of Gauss, these are very rough estimates due to limited knowledge of planetary interiors and dynamo processes (see, e.g. \citealt{ChristensenAubert2006,Christensen2009,Yadav2013,Reiners2010,Kilmetis2024,Elias-Lopez2025}).
Indirect observational estimates of HJ's magnetic fields from star-planet interactions suggested magnetic field strengths of 20 G to 120 G \citep{Scharf2010,Vidotto2010,Vidotto2011,Cauley2019,Savel2024}.

Several studies addressed magnetic drag in General Circulation Models (GCMs) using a universal drag timescale applied throughout the planetary atmosphere \cite[e.g.,][]{Komacek2016, TanKomacek2019, Carone2020} or simplified diffusivity-based descriptions: \cite{Perna2010, Perna2010b} parametrized the Lorentz force acting on the plasma with a drag timescale and estimated a uniform Rayleigh friction at each pressure level, neglecting horizontal variations of the diffusivity in the atmosphere.
The scalar drag timescale was applied as a Rayleigh friction term on the horizontal winds.
This constant Rayleigh friction term will be referred to as 'uniform drag' throughout this work.
These models do not capture the different conditions of the day- and nightside, with the nightside possibly being much less ionized and decoupled from the magnetic field and the dayside being partially coupled to the field.  
\cite{RauscherMenou2013, Beltz2022} improved this model by introducing a spatially varying but directionally isotropic drag coefficient, calculated from local thermodynamic conditions (e.g. temperature and density) and updated at each time step. 
\citet{Beltz2022} referred to this as 'active magnetic drag', a term that will be used throughout this work. 
In these models, the drag is implemented as a scalar damping term in the momentum equation and is applied only to the zonal (east–west) wind component.
In several follow-up studies, Beltz and collaborators explored the dynamical and observational consequences of magnetic drag in (U)HJ atmospheres using the spatially varying drag description in their GCM \citep{Beltz2022b, Beltz2023, Beltz2024, Beltz2025}. 
These studies extensively investigate the effect of planetary magnetic fields on atmospheric circulation, phase curves, high-resolution emission and transmission spectra for different (U)HJs, as well as the effects on HJs with eccentric orbits.
While these models have provided valuable insight into the importance of magnetic field effects, they neglect the anisotropic response of the partially ionized gas to magnetic fields.
Based on the work  of \cite{RauscherMenou2013, Beltz2022}, \cite{Christie2025} derived a drag parametrization from an explicit calculation of the Lorentz force in a dipolar field geometry and showed that including anisotropic drag in the zonal, meridional, and vertical direction (due to Hall and Pedersen currents) modifies the flow in HJ atmospheres differently compared with models that only apply zonal drag. 
Furthermore, \cite{Rogers2017} demonstrated the effects of magnetic fields on the winds
of HAT-P-7~b using spherical 3D anelastic magnetohydrodynamic (MHD) code with the same spatially varying resistivity which is linked to temperature dependent ionization as \cite{RauscherMenou2013}.  Their simulations showed wind variability in the atmosphere is linked to the magnetic field strength and coincides with the position of the hot spot in the atmosphere. Assuming that the observed variable winds on HAT-P-7~b are due to magnetism a minimum magnetic field strength of 6~G can be constrained.

In the study presented here,  the parametrized magnetic drag from the differences between the charged particle velocity and the neutral wind velocity is derived.
The magnetic drag on the atmospheric gas bulk arises from collisions between the bulk neutral flow and the ionized component in the  atmospheric gas under the influence of a  magnetic field. 
While ions and electrons experience electromagnetic forces (Lorentz forces), neutrals do not, leading to momentum transfer that can act as a drag force on the flow.
Both, Pedersen and Hall drag components are included and applied as momentum sink terms in the horizontal wind equations. 
The corresponding frictional heating is computed self-consistently from the work done by the drag force on the neutral flow.
This approach enables a more straightforward and physically motivated way to implement the magnetic drag within the GCM code \texttt{ExoRad} \citep{Carone2020,Schneider2022, Baeyens2024}.
This work adapts ionospheric  MHD formulations to the exoplanet hot Jupiter regime, where weak ionization, rapid rotation, and strong irradiation combine to produce a unique plasma environment. 
The approach of \cite{Christie2025} and the one presented in this work are based on the same underlying non-ideal MHD framework for weakly ionized atmospheres and rely on Pedersen and Hall conductivities/ drags to describe magnetic coupling in weakly ionized atmospheres.
However, they differ in their formulation and implementation way within the GCM.
\cite{Christie2025} describe the magnetic coupling as a drag on the neutral flow based on the Lorentz force, whereas here it is implemented via Pedersen and Hall drag terms derived from drift velocities between the charged particles and the neutrals.
Furthermore, in both approaches, the energy equation for the neutral gas takes into account the corresponding feedback, where the kinetic energy dissipated by the drag terms is introduced as a local heating source.
The two approaches lead to similar  parametrizations of the anisotropic drag for the parameter space considered.
The ultra-hot regime of WASP-18~b explored in this work allows the Hall term to become significant, leading to flow asymmetries that were not seen in the parameter space considered by \cite{Christie2025}.

The physical basis for separating magnetic drag into Pedersen and Hall components is well established in terrestrial and planetary ionospheric electrodynamics. 
The modeling of MHD and electrodynamic coupling between neutral atmospheres and ionospheric currents has been a central topic in the Solar System atmospheric modeling. The Thermosphere–Ionosphere–Electrodynamics General Circulation Model (TIE-GCM) \citep{Richmond1992,RichmondThayer2000} and related thermosphere/ionosphere GCMs (e.g., \citealt{Rees1989, Wang2004}) solve the coupled neutral-ion dynamics and electrodynamics self-consistently and include ion drag and frictional heating in the momentum and energy equations. 
Additionally, \cite{Zhu2005} derived expressions for the ion drag and Joule heating for the neutral atmosphere in the thermosphere   dependent on the magnetic field, the Pedersen and Hall conductivities, the ion cyclotron frequency and the ion-neutral collision frequency.
These approaches show how ionospheric currents create frictional forcing on neutral winds and localized heating in planetary upper atmospheres. Similar studies with coupled magnetosphere–ionosphere–thermosphere systems for Jupiter and Saturn demonstrate that field-aligned currents and ionospheric torques can reshape thermospheric winds and energetics, and that the spatial variations of conductivity and currents are important for the resulting drag distribution \citep{Bougher2005,Yates2020, MullerWodarg2012}.
Neither the model presented here nor the model of \cite{Christie2025} explicitly solve the charged particle dynamics or couples the atmosphere to the magnetosphere, as it was done for Earth and for solar system giant planet GCMs. 
Nevertheless, these parameterizations are crucial since they provide a computationally feasible way to capture the main effects of a planetary magnetic field on the wind structures and heat transport in the (U)HJ's atmospheres, and allow a direct comparison to observations. 
Such models are, hence,  more advanced than simplified Rayleigh drag approaches, and represent an important step toward coupled neutral–plasma GCMs for exoplanets. 
 
The study and analysis of the magnetic drag and frictional heating is presented for the atmosphere of UHJ WASP-18 b.  WASP-18 b is used as an example to present and discuss a new modelling approach that enables to consider anisotropic magnetic drag and its resulting effect on the exoplanet climate state. The ultimate aim is to prove a model that allows to derive the global magnetic field strength which is otherwise inaccessible for extrasolar planets.  WASP-18 b is an UHJ with an equilibrium temperature of 2400 K and a mass of $\sim10$ Jupiter masses. It orbits an F6-type star at a close distance of 0.02 AU \citep{Hellier2009,Southworth2010}. 
Recent Hubble Space Telescope (HST) and Transiting Exoplanet Survey Satellite (TESS) phase‑curve observations revealed an inefficient day–night heat circulation ($\gtrsim96\%$ of the heat remains on the dayside) and a small hotspot offset 3°-5° eastwards \citep{Arcangeli2019,Shporer2019}. 
Dayside thermal emission spectrum obtained with the Near-Infrared Imager and Slitless Spectrograph (NIRISS) on the James Webb Space Telescope (JWST) further exposed water emission features \citep{Coulombe2023}.
These observed features suggest the presence of an effective magnetic drag and super-solar metallicity \citep{Deline2025, Arcangeli2019} and makes it a suitable laboratory for studying magnetic effects in exoplanet atmospheres. 
The approach presented here offers a more accurate parametrization of the magnetic drag and heating, and opens the way for more accurate predictions of circulation patterns, temperature distributions, and potentially observable signatures such as phase curves and wind speeds.
Magnetic drag on UHJs not only affects winds and temperatures but also where and how clouds may form, and can be observed in these extreme atmospheres (see, e.g., \cite{Helling2021, Kennedy2025}).

This study investigates the following research questions:
\begin{itemize}
   \item  To what extent disrupts magnetic drag the equatorial superrotation of gas giants? Can it change the overall flow pattern?
   \item How does magnetic drag modify the heat redistribution between the day- and nightsides?
    \item At which gas pressures does magnetic drag become dynamically important for the atmosphere?
\end{itemize}

\noindent
The results show that the equatorial jet is weakened  by a magnetically coupled atmosphere (independent on the details of magnetic drag treatments), anisotropic drag that traces the locally changing ionization in particular may lead to a partial decoupling of the day-night flow on the morning terminator in the upper atmosphere.

The paper is organized as follows: 
Sec. \ref{sec:approach} summarizes the approach that addresses the research questions, gives an overview of the GCM code \texttt{ExoRad}, which physics is implemented,  and the physical and numerical parameters (incl. boundary layers, magnetic drag implementation)  used for the WASP-18 b simulations. 
Sec. \ref{sec:parametrized_drag} presents the analysis of the plasma parameters (incl. degree of ionization, magnetic Reynolds number) and the  detailed derivation of the anisotropic drag parametrization, which is implemented in \texttt{ExoRad}. In Secs. \ref{sec:results} and \ref{sec:discussion}, the results of the simulations with different drag treatments are shown and their influence on the wind pattern and gas temperature distribution are compared, and the limitations of the model are discussed. The effect on the climate parameters like the day-night gas temperature difference, terminator temperature difference, jet spread and width are explored. In Sec. \ref{sec:summary}, a summary of this study is presented. In the Appendix, an overview is given of  the active drag treatment based on the models of \cite{Perna2010,RauscherMenou2013} and the drag treatment of \cite{Christie2025}.
\section{Approach}\label{sec:approach}
This study investigates how magnetic drag
affects the atmospheric dynamics of UHJs that show extreme day/night-differences,
exploring the example of WASP-18 b. WASP-18 b was chosen for this study because it is one of the hottest close-in extrasolar gas giants  \citep{Parmentier2018} with an extreme irradiation environment. The high dayside temperatures reaching a maximum brightness temperature of about 3000 K between 1~bar and 10$^{-2}$~bar \citep{Coulombe2023} lead to high thermal ionization of alkali metals and other species (incl. Al, Fe, Ti; see \citealt{2019A&A...626A.133H}), resulting in noticeable electrical conductivities and thus strong magnetic coupling in the atmosphere. Furthermore, its high mass and short orbital period hint the presence of a strong planetary dipole magnetic field \citep{Elias-Lopez2025}, making it a useful test case for studying magnetic drag effects on atmospheric circulation. Our approach to answer our research question is
divided into four steps: (1) 
The atmospheric thermal structure of WASP-18 b obtained from \texttt{ExoRad} simulations is analyzed to evaluate where magnetic effects are expected to be significant. 
Therefore, the ionization fraction (Eq.~\ref{eq:xe}) is calculated from equilibrium chemistry using  \texttt{ExoRad} gas temperature and gas pressure profiles, and it is applied to evaluate magnetic coupling through plasma parameters such as magnetic Reynolds number (Eq.~\ref{eq:rm_calc}) and the electron plasma frequency (Eq.~\ref{eq:pe}). 
Furthermore, the analysis allows to identify where non-ideal MHD effects such as the Hall effect might become important (Sec. \ref{sec:discussion}).
(2) From the three-fluid description of the atmospheric gas (Eqs.~\ref{eq:neutrals}-\ref{eq:ions}), including electrons, ions, and neutral particles coupled through neutral-plasma collisions (Eqs.~\ref{eq:nuin}, \ref{eq:nuen}), a parameterization of the magnetic drag in a partially ionized atmosphere is derived that can be used in a GCM. 
The parameterization separates between Pedersen (dissipative) and Hall (non-dissipative) components, leading to an anisotropic drag (Eq.~\ref{eq:drag_final}). The derivation 
allows to couple local plasma properties (ionization fraction, collision frequencies, magnetic field strength) to momentum exchange with the neutral atmosphere. (3) The anisotropic drag parameterization is implemented into \texttt{ExoRad}. For comparison, other known magnetic drag models are summarized (Appendix \ref{ap:active_drag} and \ref{ap:christe}) and runs with no drag and with simplified drag parameterizations ('uniform' and 'active drag') are performed to analyze the consequences of different magnetic drag treatments on the atmospheric dynamics, using WASP-18 b as example. (4) Finally,  the atmospheric dynamics, wind speeds, and gas temperatures across the different drag 
treatments are compared. 
These results are analyzed further through parameters that characterize the planet's climate state as introduced by  \cite{Plaschzug2025}: the day-night and evening-morning terminator temperature  difference,  wind jet speed and width.
Additionally, a scale analysis of Ohm's law is performed to evaluate the relative importance of Pedersen and Hall currents and ambipolar diffusion at different gas pressure levels. 
This is important to determine whether the drag is isotropic or directionally dependent, and at what altitudes magnetic forces can modify the large–scale circulation.
In weakly ionized atmospheres, ions and electrons drift relative to the neutrals because only the charged species are directly affected by the magnetic field. This mechanism is usually referred to as ambipolar diffusion in astrophysics, and it allows the magnetic field to slip through neutral gas via collisions between charged and neutral particles, which transfer momentum between the plasma and the neutrals. The collisional momentum exchange can result in frictional heating and may be important for energy dissipation (e.g., \citealt{Khomenko2012, Hillier2024}).
Identifying which magnetic process dominates is  important for understanding the physical regime of the atmosphere and the effect of magnetic forces on the atmospheric flow.
\section{3D atmosphere model: \texttt{ExoRad}}
To model the 3D atmospheric hydro- and thermodynamics of WASP-18 b under different magnetic drag treatments, the 3D GCM  \texttt{ExoRad} \citep{Carone2020,Schneider2022}, is employed that uses the hydrodynamical core \texttt{MITgcm} \citep{Adcroft2004} to solve the Navier-Stokes equations for a hydrostatic atmosphere on a rotating sphere, complementing it with physical parametrizations suitable for a highly irradiated, tidally locked gas giant.
\paragraph{Model setup:}
An outline of the \texttt{MITgcm} used here is in place, given that various papers are applying the same core with their individual developments (see, e.g., \citealt{2025arXiv250921588S}).
The \texttt{MITgcm} model applied here solves the hydrostatic primitive equations (HPE, \citealt{Showman2009}) using the finite-volume method to discretize the momentum equations in space and the second-order Adams-Bashforth time integration scheme \citep{Lilly1965} for explicit time stepping on a rotating sphere in a C32 cubed-sphere grid, corresponding to a horizontal resolution of 128 $\times$ 64 in longitude ($\phi$) and latitude ($\theta$) ($2.8^\circ\times 2.8^\circ$).
The dynamical core uses a staggered Arakawa C grid \citep{ArakawaLamb1977,Showman2009}. 
The equation of state is the ideal gas law.
The vertical coordinate is defined as gas pressure $p_\mathrm{gas}$~(computed in Pa).
However, for clarity, all pressure values discussed in the text and shown in figures are expressed in bar.
The modeled atmosphere extends from  $10^{-5}$ bar to 700 bar.
Pressure ($p$-) coordinates are preferred in atmospheric modeling because they align better with hydrostatic balance.
The full set of equations (horizontal momentum,  vertical momentum under hydrostatic equilibrium, mass continuity, thermodynamic energy equations) in $p$-coordinates, which are solved in ExoRad are:
\begin{align}\label{eq:horizontalv}
    \frac{d\underline{v}_h}{dt}=-f\underline{\hat{k}}\times \underline{v}_h-\nabla_{p_\mathrm{gas}}\Phi+\underline{F}_v,\\\label{eq:vericalv}
    \frac{\partial\Phi}{\partial p_\mathrm{gas}}=-\frac{1}{\rho_n},\\
    \nabla_{p_\mathrm{gas}}\cdot\underline{v}_h+\frac{\partial\omega}{\partial p_\mathrm{gas}}=0,\\\label{eq:energyp}
    \frac{d\Theta}{dt}=\frac{\Theta}{T_\mathrm{gas}}\frac{q}{c_p},
\end{align}
with the time $t$~[s], the horizontal velocity component of the neutral gas on pressure surfaces $\underline{v}_h=(u,v,0)$, the zonal velocity $u$~[m/s], the meridional velocity $v$~[m/s], the gas temperature $T_\mathrm{gas}$~[K], the mass density of the neutral gas $\rho_n$~[kg m$^{-3}$], the horizontal gradient $\nabla_{p_\mathrm{gas}}$ evaluated on constant pressure surfaces, the vertical velocity in pressure coordinates $\omega=\frac{dp_\mathrm{gas}}{dt}$~[Pa/s], the geopotential on constant-pressure surfaces $\Phi$~[$\mathrm{m^2s^{-2}}$], the Coriolis parameter $f=2\Omega\sin\phi$~[$\mathrm{s^{-1}}$], the planetary rotation rate $\Omega$~[$\mathrm{rad\,s^{-1}}$], the local vertical unit $\underline{\hat{k}}$, the external sink term in the horizontal momentum $\underline{F}_v$, that comprises Rayleigh damping (see Eqs. \ref{eq:sponge} and \ref{eq:basal}) and frictional drag (Eq. \ref{eq:drag_final}). 
In the \texttt{MITgcm}, the gas temperature is represented by the potential temperature 
$\Theta$~[K], which is a measure of entropy\footnote{It is the temperature that a gas volume would attain under adiabatic expansion/compression if the gas would be brought to a reference pressure, which is here set to the pressure of the lower boundary equal to 700~bar.} \citep{Showman2009}:
\begin{equation}
    \Theta=T_\mathrm{gas}\left(\frac{p_0}{p_\mathrm{gas}}\right)^{R_d/c_p},
\end{equation}
where  $R_d$~[$\mathrm{J\,kg^{-1}K^{-1}}$] is the specific gas constant, $c_p$~[$\mathrm{J\,kg^{-1}K^{-1}}$] is the specific heat at constant pressure, and $p_0$~[Pa]  the reference pressure ($p_0=7\times10^7$~Pa).
The thermodynamic energy equation includes the total heating rate per unit mass $q=q_\mathrm{rad}+q_\mathrm{fric,deep}+q_\mathrm{fric}$~[$\mathrm{W\, kg^{-1}}$].
$q_\mathrm{rad}=g\frac{\partial F^\mathrm{net}}{\partial p_\mathrm{gas}}$~[$\mathrm{W\, kg^{-1}}$], where $g$~[$\mathrm{m s^{-2}}$] is gravity and $F^\mathrm{net}$~[$\mathrm{W m^{-2}}$]  the total bolometric flux (Eq. \ref{eq:Fnet}), represents radiative heating and cooling, calculated from the stellar irradiation by solving the radiative transfer equation.
$q_\mathrm{fric,deep}$~[$\mathrm{W\, kg^{-1}}$] is the frictional heating due to deep friction (Eq.~\ref{eq:qfricdeep}).
$q_\mathrm{fric}$~[$\mathrm{W\, kg^{-1}}$] accounts for frictional heating resulting from the dissipation of kinetic energy into thermal energy (Eq.~\ref{eq:qfric} for 'anisotropic drag', Eq.~\ref{eq:qfricactive} for 'active drag', and Eq.~\ref{eq:qfricuni} for the 'uniform drag' approach).
The total (or Lagrangian) derivative is here defined by $\frac{d}{dt}=\frac{\partial}{\partial t}+\underline{v}_h\cdot\nabla_{p_\mathrm{gas}}+\omega\frac{\partial}{\partial p_\mathrm{gas}}$.

The conservation of radiative energy, that is the balance between incoming stellar flux and outgoing planetary flux was tested for \texttt{ExoRad} in \citet{Schneider2022b} for the UHJ WASP-76 b assuming no interior heat flux ($T_{\rm int}=0$~K). The model was found to preserve radiative energy conservation to 99.9\% even for very long simulation times (86000 days). Kinetic energy and angular momentum was found to be conserved within 0.5\% accuracy for a simulation time of 2000~days \citep{Carone2020}.
\paragraph{Different drag treatment configurations:}
The effect of magnetic drag is explored and different parameterizations are compared:
'no drag', 'uniform drag', 'active drag', and 'anisotropic drag'. 
The 'no drag' model represents a purely hydrodynamic atmosphere, where magnetic effects are neglected and the flow evolves freely under the balance between radiation, pressure gradients, and planetary rotation.
In the 'uniform drag' model, a constant drag timescale of $\tau_\mathrm{drag}=10^{4}$~s is applied to both the horizontal momentum equation and energy equation (see e.g., \citealt{TanKomacek2019}).
This simplification represents a spatially uniform magnetic coupling between the atmospheric flow and the planetary interior.
Such uniform drag approaches are commonly used in HJ atmospheric circulation studies as a phenomenological representation of magnetic effects, often without specifying the underlying physical damping mechanism.
Using the inverted form of  Eq.~\ref{eq:dragtPe}, the effective local field strength can be estimated that is required to achieve the prescribed $\tau_\mathrm{drag}=10^4$~s, given the local gas density $\rho_n$ and the magnetic diffusivity $\eta$:  $B_{\mathrm{effective}}\sim\sqrt{\frac{\mu_0\rho_n\eta}{\tau_\mathrm{drag}}}$.
Since  $\rho_n$ and  $\eta$ vary with $p_\mathrm{gas}$, $\phi$ and $\theta$, 
$B_{\mathrm{effective}}$ is likewise spatially variable. For example,
$B_{\mathrm{effective}}=0.1$~G is obtained at $p_\mathrm{gas}=10^{-3}$~bar and $B_{\mathrm{effective}}=19$~G at $p_\mathrm{gas}=1$~bar (for $\phi=0$° and $\theta=0$°). 
Physically, the fixed uniform drag acts like a steady resistivity that slows down the horizontal flow everywhere, on both the hot, partially ionized dayside and the cooler, weakly ionized nightside. In reality, however, magnetic effects should vary strongly with temperature and ionization.
In contrast, the 'active' and 'anisotropic drag' models take into account a more realistic, spatially dependent treatment of magnetic coupling.
The drag strength depends on the local ionization fraction, $x_e$ (Fig.~\ref{fig:RM}, right), and magnetic field geometry, representing how magnetic forces interact more efficiently in hotter, more ionized regions 
(Sec.~\ref{sec:parametrized_drag} and Appendix \ref{ap:active_drag}). 
The 'active drag' was derived from the assumption that the weakly ionized atmospheric winds experience a bulk Lorentz force drag (with the detailed formulation given in Appendix~\ref{ap:active_drag}) and this effect is parameterized with a drag timescale which is described here by $\tau_\mathrm{drag}=\frac{\mu_0\rho_n\eta}{B_{\mathrm{dip},r}^2}$ with the radial component of the dipolar magnetic field $B_{\mathrm{dip},r}$ (Eq.~\ref{eq:Bdip}). The external sink term in the zonal  momentum is implemented in \texttt{ExoRad} as $F_{v,\phi}=-u/\tau_\mathrm{drag}$ and the frictional heating as $q_\mathrm{fric}=u^2/\tau_\mathrm{drag}$.
In the active drag case, the magnetic coupling acts only on the zonal winds, under the assumption that these dominate the large-scale circulation. 
However, because meridional winds can be comparable in magnitude, this simplification may miss important effects. 
The potential limitation of the active drag was discussed in \cite{RauscherMenou2013} and was investigated further in \cite{Christie2025}.
The anisotropic drag model, which is described in Sec.~\ref{ch:momentum_exchange}, includes drag in zonal and meridional directions, creating a more complete representation of how magnetic drag modifies the flow across field lines and determines the global wind pattern.
In the anisotropic drag approach, the drag is decomposed into Pedersen and Hall components, resulting in the drag force that both damps and redirects the flow relative to the magnetic field geometry
\paragraph{Numerical parameters:}
The vertical space covers 47 vertical cells.
A logarithmic spacing with 41 grid cells is applied between $10^{-5}$ bar and 100 bar to resolve the radiative and dynamically active regions, while linear spacing with six grid cells is used between 100 bar and 700 bar with a step size of 100 bar.
The vertical resolution of the model resolves the atmosphere with about two vertical grid layers per scale height, which is important for accurately capturing vertical dynamics.
The dynamical time step $\Delta t=$25~s is chosen in the model to satisfy both  the Courant–Friedrichs–Lewy (CFL) condition (set by horizontal advection and grid spacing) and  the shortest physical timescales introduced by parameterized source terms.
Furthermore, a fourth-order Shapiro filter \citep{Shapiro1970} with a dampening time scale of 25~s is applied which horizontally smooths grid-scale noise \citep{Showman2009, Carone2020}.
Although the application of smoothing methods is routinely used in  GCMs, they still might have a non-negligible effect on the atmospheric dynamics (e.g., \citealt{Heng2011,Polichtchouk2014,Skinner2021}).
The simulations ran for a total of 1000 planetary days to allow the temperature structure to stabilize and the system to approach a steady-state solution.
\paragraph{Boundary conditions:}
The top boundary of the computational domain acts like a solid boundary ($\omega=0$~Pa/s),
while the bottom boundary is an impermeable surface placed at $p_\mathrm{bottom}=p_0=700$~bar located below the region of interest.
At both the top and bottom boundaries, a free slip boundary condition is applied to the horizontal velocity.
To dampen nonphysical gravity wave reflection at the top of the computational volume, a sponge layer (ghost cells) is applied between $p_\mathrm{gas}=10^{-4}$~bar and $p_\mathrm{gas}=10^{-5}$~bar \citep{Carone2020,Beltz2022, Schneider2022}.
Within this layer, Rayleigh friction is applied to the  zonal velocity $u$, which acts to bring $u$ closer to its longitudinal mean $\bar{u}$~[m/s] via
\begin{equation}\label{eq:sponge}
    \frac{du}{dt}=-\tilde{k}(u-\bar{u})
\end{equation}
where the strength parameter $\tilde{k}$ [s$^{-1}$] is a function of $p_\mathrm{gas}$:
\begin{equation}
    \tilde{k}(p_\mathrm{gas})= k_\mathrm{top}\,\mathrm{max}\left[0, \,1-\frac{p_\mathrm{gas}^2}{p_\mathrm{sponge}^2}\right]^2.
\end{equation}
The parameters $k_\mathrm{top}$~[s$^{-1}$] and $p_\mathrm{sponge}$~[Pa] determine the intensity and the location of the Rayleigh friction in the sponge layer.
The default values in this model are $k_\mathrm{top}=1728\times10^3$~s$^{-1}$ and $p_\mathrm{sponge}=10$~Pa \citep{Deline2025}.
Since the domain uses a cubed sphere grid, Eq.~\ref{eq:sponge} is computed  by first converting the cubed sphere grid values to geographic coordinates, averaging the deprojected $u$  within 20 latitude bands, and finally mapping the resulting $\bar{u}$ back onto the cubed sphere grid \citep{Schneider2022}. 
This upper sponge layer isolates potentially nonphysical solutions caused by unresolved boundary effects from the interior test volume 
(see, e.g.,  also Sec.~2.3. in \citealt{2004A&A...423..657H}).
Consequently, the numerical solution in the  region $p_\mathrm{gas}<10^{-4}$~bar is excluded from  the analysis of the atmospheric dynamics presented in Sec.~\ref{sec:results}.
Similarly, a deep-layer sponge drag is applied at the bottom boundary to ensure numerical stability, following \cite{Carone2020}. 
With this method shear flow instabilities and nonphysical changes in the flow pattern of the simulation domain are avoided without directly affecting the observable atmosphere because the drag only modifies the winds in the deepest layers ($p_\mathrm{gas}>490$~bar).
This deep-layer sponge drag dissipates the horizontal wind $\underline{v}_h$ via
\begin{equation}\label{eq:basal}
    \frac{d\underline{v}_h}{dt}=-k_\mathrm{deep}\underline{v}_h.
\end{equation}
The parameter $k_\mathrm{deep}$~[s$^{-1}$] is defined by
\begin{equation}
    k_\mathrm{deep}=k_\mathrm{bottom}\, \mathrm{max}\left[0,\,\frac{p_\mathrm{gas}-4.9\times10^7~\mathrm{Pa}}{p_0-4.9\times10^7~\mathrm{Pa}}\right]
\end{equation}
with the control parameter $k_\mathrm{bottom}=86400$~s$^{-1}$, 
which dissipates fast wind jets at depth \citep{Carone2020}. The kinetic energy dissipated at the bottom via friction is converted into thermal energy, approximating Ohmic dissipation  \citep{RauscherMenou2013,Carone2020}. The corresponding heating rate  per unit mass of the neutral gas is given by
\begin{equation}\label{eq:qfricdeep}
    q_\mathrm{fric,deep}=k_{\rm deep} \underline{v}_h^2. 
\end{equation}
Simulation tests for the sponge layer are discussed in detail in \citet{Carone2020}.
\paragraph{Atmospheric initialization:}
The atmosphere was initialized using the analytic temperature profile of \cite{Parmentier2015}. 
Therefore, the interior and equilibrium temperatures are used.
The equilibrium temperature for zero albedo,
\begin{equation}\label{eq:eqT}
    T_\mathrm{eq,0}=T_\star~\mathrm{[K]}\sqrt{\frac{R_\star~\mathrm{[m]}}{2 a_p~\mathrm{[m]}}},
\end{equation}
was computed from stellar and orbital parameters (see Tab.~\ref{tab:wasp18b_params}). 
For WASP-18 b, adopting stellar effective temperature $T_\star=6400$~K, stellar radius $R_\star=1.23 R_\odot$, and semi major axis $a_p=0.0203$~AU  yields $T_\mathrm{eq,0}\approx 2402$~K. 
$T_\mathrm{eq,0}$ was then applied to estimate the corresponding interior temperature of $T_\mathrm{int}\approx 573$~K following \cite{Thorngren2019}. 
These temperatures define the initial radiative–convective profile using the analytical model of \cite{Parmentier2015}, which yields a hot adiabat  potential temperature of $\Theta_\mathrm{ad}=6940$~K  at 1 bar.
Synchronous rotation and a surface gravity of $g=190~\mathrm{m\,s^{-2}}$ are assumed. This setup provides a hot, physically consistent initial state in the deep layers.
\cite{SainsburyMartinez2019} have shown that the time required for the deep atmosphere to reach equilibrium after cooling from a hot initial condition is less than the time required for heating it from a cold initial state to a hotter one.
\paragraph{Radiative transfer treatment:}
The \texttt{ExoRad} framework uses the full radiative transfer treatment expeRT/MITgcm \citep{Carone2020,Schneider2022} based on  \texttt{petitRADTRANS} \citep{Molliere2019, Molliere2020}.
The implementation of radiative transfer in \texttt{ExoRad} is described in detail in Sec.~2.2 of \cite{Schneider2022}.
The model solves the radiative transfer equation
\begin{equation}\label{eq:radiative}
    \underline{n}\cdot\nabla I_\nu =\alpha^\mathrm{tot}_\nu(S_\nu-I_\nu),
\end{equation}
with the unit vector $\underline{n}$, the intensity $I_\nu$~[$\mathrm{Wm^{-2}sr^{-1}Hz^{-1}}$], the frequency $\nu$~[Hz], the source function $S_\nu$~[$\mathrm{Wm^{-2}sr^{-1}Hz^{-1}}$], and the inverse mean-free path of the light beam $\alpha^\mathrm{tot}_\nu$~[m$^{-1}$].
Eq.~\ref{eq:radiative} describes how a beam of radiation traveling through a planetary atmosphere loses energy through absorption, gains energy through emission, and redistributes energy through scattering.
The equation is solved for photons originating in the planetary atmosphere or scattered from incoming stellar radiation. 
Direct stellar intensity extinction is modeled separately using exponential decay, and the total intensity is obtained by summing all contributions (intensities of planetary and scattered stellar photons, and attenuated stellar intensity).
The mean stellar attenuated intensity in a plane-parallel atmosphere is 
\begin{equation}
    J^\star_\nu=\frac{I^\star_\nu(p_\mathrm{gas}=p_\mathrm{top})}{4\pi}\exp \left(\frac{-\tau_\nu}{\mu_\star}\right),
\end{equation}
with the stellar intensity at the top of the atmosphere $I^\star_\nu(p_\mathrm{gas}=p_\mathrm{top})$~[$\mathrm{W m^{-2}sr^{-1}Hz^{-1}}$], calculated using the PHOENIX stellar model spectrum \citep{Husser2013} via \texttt{petitRADTRANS}. 
For a tidally locked planet, the cosine of the zenith angle, that is, the angle between the normal vector on top of a given atmospheric column on the planet and the incoming stellar light is $\mu_\star=\cos\theta \cos\phi$, with latitude $\theta$ and longitude $\phi$, when the origin of the coordinate system is set to the substellar point. 
The optical depth is
\begin{equation}
    \tau_\nu=\int\frac{\kappa^\mathrm{tot}_\nu}{g}dp_\mathrm{gas},
\end{equation}
where $\kappa^\mathrm{tot}_\nu=\alpha^\mathrm{tot}_\nu\rho_n$~[$\mathrm{m^2kg^{-1}}$] is the total gas-phase opacity given from absorption and scattering coefficients.
The received bolometric stellar flux is computed via 
\begin{equation}
    F^\star=4\pi\mu_\star\int_\nu J^\star_\nu d\nu~[\mathrm{Wm^{-2}}].
\end{equation}
The radiative transfer equation (Eq. \ref{eq:radiative}) was solved using the Feautrier method \citep{Feautrier1964} with accelerated lambda iteration \citep{Olson1986} to obtain the planetary flux and 
iterated to convergence of the planetary intensity field.
Convergence is achieved when the relative change in the source function falls below 2 \% of the reciprocal local lambda operator, providing a good balance between accuracy and performance \citep{Schneider2022}.
The previous radiative time-step’s source function serves as the initial guess, allowing later time-steps to converge in only a few iterations (less than three), compared to 50–100 iterations in the beginning (first 100 time steps).

Then the bolometric flux $F^\mathrm{pla}$~[$\mathrm{W m^{-2}}$] was computed by integrating the radiation field $F^\mathrm{pla}_\nu$:
\begin{equation}
    F^\mathrm{pla}=\int_\nu F^\mathrm{pla}_\nu d\nu =-4\pi\int_\nu H^\mathrm{pla}_\nu d\nu.
\end{equation} 
$H^\mathrm{pla}_\nu$~[$\mathrm{W m^{-2}Hz^{-1}}$] is the directional flux of radiation at frequency $\nu$ emerging from the planet.
The total bolometric flux
$F^\mathrm{net}$~[$\mathrm{W m^{-2}}$]  was then obtained as 
\begin{equation}\label{eq:Fnet}
    F^\mathrm{net}=F^\mathrm{pla}+F^\star.
\end{equation}
Following \cite{Showman2009}, $F^\mathrm{net}$
was evaluated on vertically staggered cell interfaces, using quadratic Bézier interpolation for temperature profiles \citep{Lee2022} to enhance accuracy and stability in particular at the terminators, where the zenith angle of incoming radiation is approaching zero. 
$F^\mathrm{net}$ was computed for every second grid column, with linear interpolation between columns and direct computation at tile borders due to the C32 grid geometry.

The total gas-phase opacity $\kappa^\mathrm{tot}_\nu$ is represented by correlated-k tabulated line gas opacities\footnote{For most molecules, opacities are taken from the ExoMol data base \citep{Tennyson2016_ExoMol,TennysonEtal2020jqsrtExomol2020}.} resolved into 11 wavelength bins from 0.26~$\mathrm{\mu m}$ to 300~$\mathrm{\mu m}$ \citep{Kataria2013,Schneider2022} corresponding to the S1 resolution for the following species: 
$\mathrm{H_2O}$ \citep{H2OBarber2006}, Na \& K  including pressure broadening \citep{Allard2019}, $\mathrm{CO_2}$ \citep{CO2Yurchenko2020}, $\mathrm{CH_4}$ \citep{CH4Yurchenko2017}, $\mathrm{NH_3}$ \citep{NH3Coles2019}, CO \citep{COLi2015},  $\mathrm{H_2S}$ \citep{H2SAzzam2016}, HCN \citep{HCNBarber2014}, SiO \citep{SiOBarton2013}, $\mathrm{PH_3}$ \citep{Khomenko2014}, FeH \citep{FeHWende2010}, TiO \citep{TiOMcKemmish2019} and VO \citep{VOMcKemmish2016}, as well as  $\mathrm{H^-}$ scattering suitable for an ionized atmosphere (see Tab.~1 in \citealt{Schneider2022}).
The gas line opacities are pre-calculated on a grid of gas pressure and temperature ($T_{\rm gas}=100$~K $\ldots 4000$~K with linear spacing $\Delta T=3.9$~K  and $p_{\rm gas} =700$~bar $\ldots 10^{-5}$~bar as set by the vertical atmospheric pressure grid), assuming local chemical equilibrium using \texttt{easyCHEM} \citep{Lei2025}.
Furthermore, the opacities also include collision-induced absorption from H$_2$-H$_2$ \citep{Borysow2001, Borysow2002, Richard2012} and H$_2$-He  \citep{Borysow1988, BorysowFrommhold1989, Borysow1989}, Rayleigh scattering by H$_2$ \citep{DalgarnoWilliams1962} and He \citep{ChanDalgarno1965}, and opacities for H$^-$ free-free and bound-free photon absorption in the presence of free electrons \citep{Gray2008,Jacobs2022}. Atmospheric chemistry assumes solar metallicity ([Fe/H] = 0) and a solar C/O ratio of 0.55 under local thermodynamic equilibrium (LTE) conditions \citep{Asplund2009}.

As the upper ($p_\mathrm{gas}<1$~bar) atmospheric temperatures can evolve significantly, radiative fluxes are recalculated every fourth dynamical time step, that is, a radiative time step of $\tau_\mathrm{rad}=100$~s is used.
\paragraph{Planetary system properties:}
For the equatorial magnetic field strength at the reference radius $R_\mathrm{ref}$~[m], $B_0=5$ G is chosen  (\citealt{Coulombe2023}).
\cite{Coulombe2023} showed that an internal magnetic field of at least 5 G is  required within the framework of a GCM with active magnetic drag \citep{Beltz2022} to  reproduce the observed white-light curve, as such a field effectively suppresses any noticeable longitudinal shift of the hotspot from the substellar point.
However, this threshold is likely model dependent and may vary with the adopted drag formulation and atmospheric conductivity profile.

For $R_\mathrm{ref}$, the transit radius of $R_\mathrm{p,obs}=1.19\, R_\mathrm{Jup}$ (Tab.~\ref{tab:wasp18b_params}) is adopted which is derived from optical light curves corresponding to the altitude where the slant optical depth reaches unity (e.g., \citealt{Fortney2005}).
This contrasts with Jupiter, where the radius is conventionally defined at 
$p_\mathrm{gas}=1$ bar \citep{Seidelmann2007}.
The optical transit radius is wavelength-dependent and sensitive to clouds and hazes affecting atmospheric opacities and possibly  leading to an overestimated radius (see, e.g., \citealt{Burrows2007}). 
Unlike at Jupiter, there is no fixed physical pressure level for the radius across different exoplanets due to their large diversity.
However, the transit radius is the  directly observable radius and is consistent with the observed transit depth and the derived planetary mass–radius relation.
Using representative parameters for WASP-18~b ($R_d=4343\,\mathrm{J\,kg^{-1}K^{-1}}$, $g=190\,\mathrm{m\,s^{-2}}$, see Tab.~\ref{tab:wasp18b_params}) and the $T_\mathrm{gas}(p_\mathrm{gas})$ profile at $\phi=0$° and $\theta=0$° from \texttt{ExoRad}, the  geometric height difference $\Delta z$ can be calculated by integrating the hydrostatic equation
$dz=-\frac{R_dT_\mathrm{gas}(p_\mathrm{gas})}{g} d\ln p_\mathrm{gas}$ between $p_\mathrm{obs}=10^{-3}$~bar and $p_\mathrm{target}=1$~bar.
Mapping $R_\mathrm{p,obs}$ at $p_\mathrm{obs}=10^{-3}$~bar to $p_\mathrm{target}=1$~bar gives $\Delta z\approx467$~km which results in less than 0.6~\% of $R_\mathrm{p,obs}$.
The overview of the simulation parameters and the setup of the different drag treatments in the performed simulations is given in Tables~\ref{tab:wasp18b_params} and \ref{tab:drag_params}.
\paragraph{Numerical stability and drag implementations:}In all simulations with active or anisotropic magnetic drag, it was  found that numerical stability imposes a lower limit on the drag timescale $\tau_\mathrm{drag}$~[s]. 
The drag timescale is defined by $\tau_\mathrm{drag}=\min(\rho_n/K_H\,, \rho_n/K_P)$, where  $\rho_n$~[$\mathrm{kg\, m^{-3}}$] is the gas density, and $K_p$~[$\mathrm{kg\,m^{-3}s^{-1}}$] and $K_H$[$\mathrm{kg\,m^{-3}s^{-1}}$] are the total Pedersen and Hall drag coefficients
(Eqs.~\ref{eq:totalKH}, \ref{eq:totalKP}).
Specifically, when $\tau_\mathrm{drag}<100$~s, the model becomes unstable and crashes due to the strong damping in the low-pressure upper atmosphere. 
Such short timescales are  often reached on the dayside at  $p_\mathrm{gas}<10^{-2}$~bar, where ionization is high and diffusivity is low (gray line in Fig. \ref{Fig:drag_coeff}). 
To maintain numerical stability across the simulation runs,  a global lower bound of $\tau_\mathrm{drag}\geq 100$ s is imposed which corresponds to four times the model time step ($\Delta t=25$~s).
Similar lower bounds of $\tau_\mathrm{drag}$ are commonly used in GCMs with active drag (e.g., \citealt{RauscherMenou2013, Coulombe2023}).
\begin{table*}[htbp]
\centering
\caption{Physical (top) and numerical (bottom) parameters for WASP-18 b \texttt{ExoRad} simulations}
\label{tab:wasp18b_params}
\begin{tabular}{lll}
\hline
\textbf{Parameter} & \textbf{Value} &\textbf{Source}\\
\hline
Planetary mass, $M_p$ & $10.4~M_{\mathrm{Jup}}$ & \citealt{Southworth2009}\\
Planetary radius$^{a}$, $R_\mathrm{p,obs}$ & $1.19~R_{\mathrm{Jup}}$ &\citealt{Shporer2019} \\
Surface gravity, $g$ & $190~\mathrm{m~s^{-2}}$& \citealt{Southworth2009} \\
Orbital period (synchronous)& $0.94~\mathrm{days}$ &\citealt{Hellier2009} \\
Stellar effective temperature, $T_*$ & $6400~\mathrm{K}$ &\citealt{Hellier2009}\\
Stellar radius$^{a}$, $R_*$ & $1.23~R_\odot$ &\citealt{Shporer2019}\\
Semi-major axis, $a_P$ & $0.02047~\mathrm{AU}$ & \citealt{Southworth2009} \\
Equilibrium temperature for zero albedo, $T_{\mathrm{eq,0}}$ & $2402~\mathrm{K}$ & Eq.~\ref{eq:eqT} \\
Metallicity [Fe/H] & 0 (solar) & \citealt{Southworth2009} \\
Mean molecular weight, $\mu$ & $1.9$& \texttt{petitRADTRANS} \\
Specific heat$^{b}$, $c_p$ & 3.04$\times10^4~\!\mathrm{J~\!kg^{-1}~\!K^{-1}}$ & \texttt{petitRADTRANS}\\
Specific gas constant, $R_d$ & $4341~\mathrm{J~kg^{-1}~K^{-1}}$ & $R_d=k_B/(\mu m_p)$\\
\hline
Pressure range & $10^{-5}\,\ldots\,700~\mathrm{bar}$ \\
Grid resolution & C32 ($\sim128 \times 64$) \\
Vertical levels & 47 \\
Equatorial surface magnetic field strength, $B_{0}$ & 5 G \\
Magnetic field geometry & Anti-aligned dipole \\
Sponge layer Rayleigh friction ($k_\mathrm{top}$)& 20 days$^{-1}$\\
Timescale for Rayleigh friction in deep layers ($k_\mathrm{bottom}^{-1}$)& 1 day \\
Radiative time step & 100 s\\
Simulation time step & 25 s\\
Simulation time length & 1000 days\\
\hline
\end{tabular}
\tablefoot{$R_\odot$ 
$\sim$6.95508 $\times$ 10$^8$ m: solar radius; $R_\mathrm{Jup}\sim$7.1492 $\times$ 10$^7$ m: Jupiter radius (equatorial, at 1~bar); $M_\mathrm{Jup}\sim$1.898 $\times$ 10$^{27}$~kg: Jupiter's mass; $m_p\sim 1.673\times 10^{-27}$~kg: mass of a hydrogen atom. \\
\tablefoottext{a}{Values were inferred from transit light curves of TESS observed in a wavelength range from 600 to 1000~nm.}
\tablefoottext{b}{$c_P$ is derived using the adiabatic temperature gradient, which was calculated using eqs. (2.50), (2.59), (2.75) of \citealt{Gordon1994}.}
}
\end{table*}

\begin{table*}[htbp]
\centering
\caption{Drag scheme configurations in WASP-18b simulations}
\label{tab:drag_params}
\begin{tabular}{lccc}
\hline
\textbf{Drag Type} & \textbf{Parameterization} & 
\textbf{Magnetic Field Strength} &
\textbf{Direction Dependence}\\
\hline
No drag & - & - & -\\
Uniform drag & $\tau_{\mathrm{drag}} = 10^{4}\,\mathrm{s}$  & -& zonal + meridional \\
Active drag & $\tau_{\mathrm{drag}}(p_\mathrm{gas}, T_\mathrm{gas},x_e)$  & 5 G & zonal\\
Anisotropic drag & $\tau_{\mathrm{drag}}(p_\mathrm{gas}, T_\mathrm{gas},x_e)$  & 5 G& zonal + meridional \\
\hline
\end{tabular}
\end{table*}
\section{Modeling the magnetic drag in the changing ionization environments of exoplanet atmospheres}\label{sec:parametrized_drag}
\begin{figure*}\label{fig:RM}
\centering
\includegraphics[width=1\textwidth]{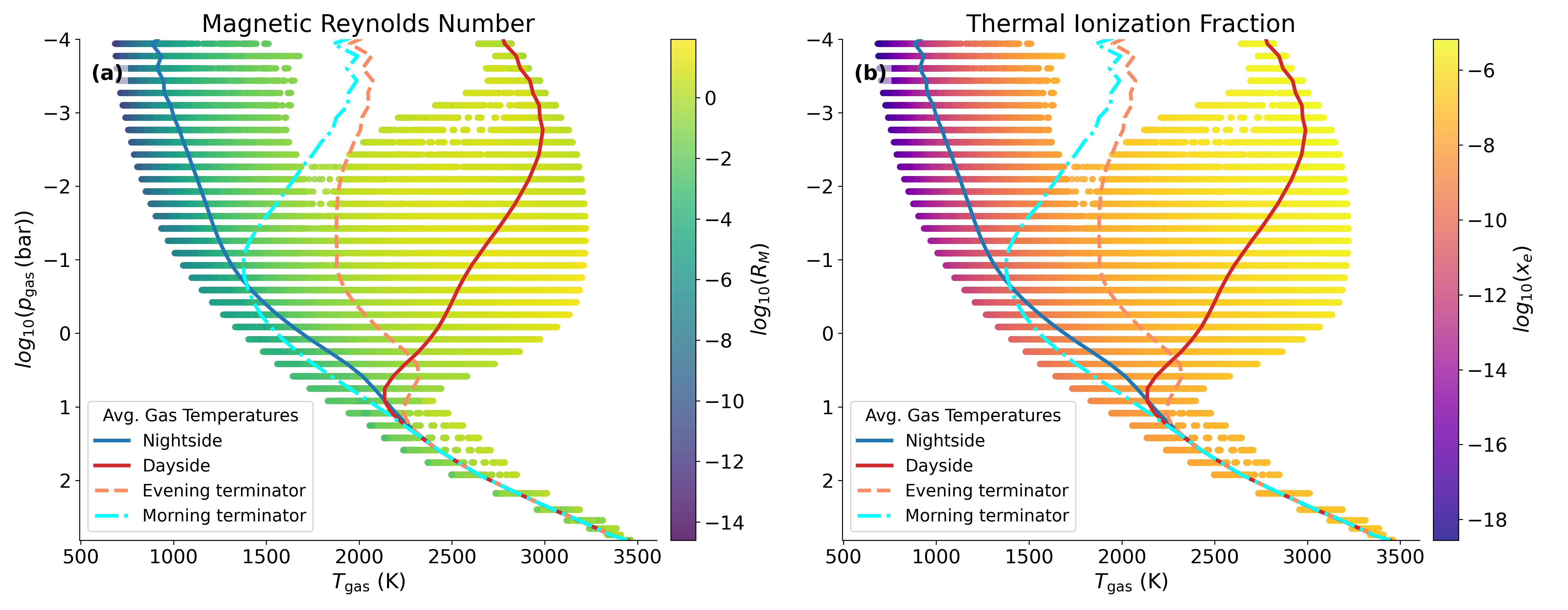}
\caption{(a) Spatial variations in magnetic Reynolds number ($R_M$ (Eq.~\ref{eq:rm_calc}), left) and (b) thermal ionization fraction ($x_e$, right) for WASP-18 b.
Large areas of the dayside are in the intermediate and high-$R_M$ regimes, while ionization varies over many orders of magnitude, motivating an anisotropic drag treatment that accounts for local conductivity and magnetic geometry. The lines show gas temperature-pressure profiles averaged over all latitudes and over different regions in the atmosphere: dayside ($-90$°$<\phi\leq90$°), nightside ($|\phi|>90$°), morning ($-97.5$°$\leq\phi\leq-82.5$°) and evening terminator ($82.5$°$\geq\phi\geq97.5$°). Results of the simulation with anisotropic drag were used for the calculation of $R_M$, $x_e$, and the averaged temperature-pressure profiles shown here.}
\label{fig:rm_xe}
\end{figure*}
The differing stellar irradiation between the dayside and nightside of WASP-18 b introduces a strong horizontal pressure gradient in the upper atmosphere. 
This gradient drives fast zonal winds from the hot to the cold hemispheres, with speeds reaching several kilometers per second. 
High temperatures in the dayside atmosphere lead to  thermal ionization of alkali metals (e.g., Na, K) and other species (e.g., Al, Ti; \citealt{2019A&A...626A.133H}) which create a partially-ionized plasma.
Figure~\ref{fig:rm_xe}b illustrates the  thermal ionization fraction $x_e$ for WASP-18 b based on the 3D \texttt{ExoRad} results.
\cite{RodriguezBarrera2015} consider $x_e>10^{-7}$ as a threshold when the gas starts to experience plasma behavior. 
The calculated $x_e$ in Fig.~\ref{fig:rm_xe}b
indicates that significant ionization already  
for upper atmospheric regions where $p_{\rm gas}<1$~bar.
Furthermore, the electron plasma frequency exceeds the electron-neutral collision frequency on the dayside for $p_{\rm gas}<0.1$~bar (dotted blue line in Fig. \ref{Fig:k_parameter}) indicating that the ionized medium behaves as a plasma in this area (e.g., \citealt{Baumjohann1996}). 
In the partially ionized upper atmosphere, frictional heating may arise from collisions between the charged particles (ions, electrons) and the dominant neutrals due to their relative drift velocity.

The importance of the effect of anisotropic drag on the atmospheric flows of (U)HJs has also been recently highlighted by \cite{Christie2025}. Their work, in which the anisotropic drag is directly derived  from the generalized Ohm's law (see Appendix \ref{ap:christe}), similarly demonstrates that the Hall effect can alter global atmospheric dynamics. 
Our work provides an alternative derivation starting from the fundamental equations of motion for the coupled ion, electron, and neutral fluids. 
This approach, which is based on the framework established for studying Earth's ionosphere-thermosphere system (e.g., \citealt{Rees1989, Song2001,Zhu2005,SchunkNagy2009}), explicitly parametrizes the drag in terms of the collision frequencies and gyrofrequencies of the existent particle species. 
This formulation represents an alternative presentation of the same underlying physics that emphasizes the role of anisotropic, collisionally driven transport and provides a physically motivated drag parametrization suitable for the implementation in GCMs.

In this section, different magnetic coupling regimes are discussed  through the magnetic Reynolds number and a parametrized form of the anisotropic magnetic drag suitable for implementation in \texttt{ExoRad} is derived.
In \texttt{ExoRad}, similar to many other GCMs (\texttt{SPARC/MITgcm} \citep{Showman2009}, \texttt{THOR} (e.g., \citealt{Mendonca2016}), \texttt{RM-GCM} \citep{RauscherMenou2010}, the Met Office Unified Model (\texttt{UM}) (e.g., \citealt{Wood2014,Edwards1996,Drummond2018})),  only the neutral fluid dynamics is evolved and the momentum equations for charged species is not solved explicitly. Instead, an additional source term representing the coupling effect through locally changing ionization is derived.

Before the model is presented, the key plasma parameters that are relevant for the characterization of the magnetic coupling are provided here calculated from the simulation without magnetic drag for the benefit of the reader.
The electron plasma frequency $\omega_{pe}$~[Hz] describes the rate at which electrons oscillate and is  given by
\begin{equation}\label{eq:pe}
    \omega_{pe}=\sqrt{\frac{n_e e^2}{m_e\epsilon_0}},
\end{equation}
with the vacuum permittivity $\epsilon_0$~$\mathrm{[C^2kg^{-1}m^{-3}s^{2}]}$, the elementary charge $e$~[C] and the electron number density $n_e$~[m$^{-3}$]. 
$\omega_{pe}$ ranges from $5\times10^{6}\,\ldots\, 9\times10^{11}$~Hz on the dayside atmosphere of WASP-18~b.
The relation $\omega_{pe}/\nu_{en}$   between  $\omega_{pe}$ and the electron–neutral collision frequency $\nu_{en}$ (Eq.~\ref{eq:nuen}) describes whether the plasma behaves more like a collisionless fluid ($\omega_{pe}/\nu_{en}>1$)
or a collisional fluid ($\omega_{pe}/\nu_{en}<1$). 
It ranges from $6\times10^{-4}$ to 50 in the dayside atmosphere.
The gyrofrequency $\omega_{cs}$~[Hz] gives the rate at which charged particles (index $s=i$ for ions and $s=e$ for electrons) gyrate around magnetic fields and is given by
\begin{equation}
    \omega_{cs}=\frac{|q_s||\underline{B}_\mathrm{dip}|}{m_s},
\end{equation}
with $q_s$~[As] the charge of species $s$, $m_s$~[kg] the mass of species $s$, and $|\underline{B}_\mathrm{dip}|$ the magnitude of the dipolar field (Eq.~\ref{eq:dipole}).
$\omega_{ce}$ ranges from $9\times10^{7}\,\ldots\,2\times10^{8}$~Hz and $\omega_{ci}$ from $1.5\times10^{3}\,\ldots\,3\times10^{3}$~Hz.
The magnetization parameter $k_s$ \citep{Leake2014} (also called Hall parameter) describes the degree to which a charged species is magnetized and is given by:
\begin{equation}\label{eq:magnetization}
    k_s=\frac{\omega_{cs}}{\nu_{sn}},
\end{equation}
with $\nu_{sn}$ the collision frequency between species $s$ and neutrals (Eqs.~\ref{eq:nuin}, \ref{eq:nuen}).
$k_i$ ranges from $2\times10^{-9}$ to $2\times10^{-2}$ and $k_e$  from $4\times10^{-7}$ to 3.8 on the dayside atmosphere of WASP-18~b.
The efficiency of magnetic induction relative to diffusion is quantified by the magnetic Reynolds number $R_M$ shown in Fig. \ref{fig:rm_xe}a. 
$R_M$ compares the timescale of magnetic field advection by the flow to the timescale of magnetic diffusion (see Eq.~\ref{eq:reynolds}) and is specifically adopted here as 
\begin{equation}\label{eq:rm_calc}
    R_M=|\underline{v}_\mathrm{h}| H_{p}\mu_0\sigma_P,
\end{equation}
with the pressure scale height $H_p=\frac{R_dT_\mathrm{gas}}{g}$~[m], the magnetic permeability $\mu_0$[$\frac{N}{A^2}$], and the Pedersen conductivity $\sigma_P$~[S/m] (Eq.~\ref{eq:hall_pedersen}).

The ionization fraction $x_e$ quantifies the proportion of particles that are ionized compared to the total number of particles in the gas and is defined as 
\begin{equation}\label{eq:xe}
    x_e=\frac{p_\mathrm{el}}{p_\mathrm{el}+p_\mathrm{gas}}\approx \frac{n_e}{n_n},
\end{equation}
with the gas and electron pressure $p_\mathrm{gas}$~[Pa] and $p_\mathrm{el}$~[Pa] and assuming that the electron temperature $T_e\approx T_\mathrm{gas}$, $p_\mathrm{gas}\gg p_\mathrm{el}$ and quasi-neutrality ($n_e\approx n_i$).
In the 3D GCM model it is assumed that all species (electrons, ions, and neutrals) have the same temperature.
The electron pressure $p_\mathrm{el}$ is calculated with the code \texttt{GGchem} \citep{Woitke2018}.
\texttt{GGchem} calculates the gas-phase chemical equilibrium using elemental abundances, local gas temperature and gas pressure to determine the number densities of all atomic and molecular species, including their ions and the electron pressure.
The ionization fraction is then interpolated to the gas pressure-temperature profile in \texttt{ExoRad} for further use in computing magnetic drag forces at each grid point in the simulation. This is faster than re-running \texttt{GGchem}.
\subsection{Where does magnetic drag affect atmosphere environments?}
The efficiency and geometry of the momentum exchange strongly depend  on $x_e$, the local magnetic field orientation, and the relative importance of inductive and diffusive processes. 
The existing 'active drag' parameterization used in GCM studies of HJs (e.g., \citealt{RauscherMenou2013}) scales with the local magnetic diffusivity $\eta$ (Eq.~\ref{eq:eta}).
While this approach captures the braking of the wind flow on the dayside, it treats the drag as isotropic only affecting the zonal component of the wind velocity but taking into account the variations of the ionization on the day- and nightside.
This approach is valid only for $R_M\ll1$ (e.g., \citealt{Dietrich2022}), where the plasma is only weakly coupled to the field, the magnetic diffusion dominates and the field lines move freely through the gas. In this case the magnetic field does only weakly affect the flow.
However, this model does not take into account that the interaction between a conducting fluid and a magnetic field is anisotropic which becomes important for the intermediate $R_M$ regime ($R_M\approx 1$).
Charged particles are constrained to move along magnetic field lines, while their movement perpendicular to the field is restricted.
This leads to an anisotropic  drag force on the neutral gas.
In Fig. \ref{fig:rm_xe}a, a map of the Reynolds number
for the modeled atmospheric structure shows 
that large regions of the dayside extend into the intermediate to high $R_M$ regime.
Inductive effects become important where $R_M> 1$  and drag is anisotropic, while the cooler nightside and deeper layers have $R_M\ll 1$. 
The anisotropic magnetic drag is represented by separating the Pedersen and Hall contributions, which are the dissipative and non-dissipative components of the  magnetic drag, respectively.
The Pedersen component corresponds to drag perpendicular to the magnetic field, leading to flow damping and kinetic energy loss, whereas the Hall component corresponds to drag perpendicular to both the magnetic field and the flow velocity modifying the flow direction without energy dissipation.
As the starting point the momentum equations for ions, electrons, and neutrals are taken (Eqs.~\ref{eq:neutrals}-\ref{eq:ions}).
Then drift velocities of the charged particles with respect to the neutrals are derived (Eq.~\ref{ref:vel_s}) which are used to parametrize the magnetic drag force (Eq.~\ref{eq:drag_final}) acting on the neutral gas.
This formulation of the anisotropic drag is applicable across low and intermediate $R_M$ regimes and is a significant physical improvement over isotropic drag models, providing a more realistic representation of the directional property of the Lorentz force in the large areas of the atmosphere where the low and intermediate $R_M$ approximation is valid.
In the region where  $R_M\gg 1$, the magnetic field lines are frozen into the plasma and advection dominates over diffusion. Here, the atmospheric gas and magnetic field are fully coupled: the gas and field move together, inducing strong currents and feedback reactions on the flow \cite{Rogers2017}. For this, a fully self-consistent MHD treatment would be required for accuracy.
However, it should be noted that $R_M$ must be interpreted carefully in the presence of strong magnetic drag. Strong drag can suppress the horizontal flow, resulting in small values of $R_M$ that represent a reduced velocity rather than weak magnetic coupling. Therefore, a small $R_M$ does not necessarily demonstrate if a magnetic drag parametrization is valid in this regime, as the flow might be in the ideal MHD limit. To analyze the system properly, not only $R_M$ but the plasma parameters introduced before need to be considered (Eqs. \ref{eq:pe}--\ref{eq:xe}).
\subsection{Momentum exchange between charged and neutral species and frictional heating}\label{ch:momentum_exchange}
The following derivations are described in a right-handed spherical coordinate system  defined by the radial distance $r$~[m], latitude $\theta$~[°], and longitude $\phi$~[°] with corresponding unit vectors $\underline{\hat{e}}_r$ (radially outward), $\underline{\hat{e}}_\theta$ (positive northward), and $\underline{\hat{e}}_\phi$ (positive eastward).
This coordinate system, which is in height coordinates, differs from the $p$-coordinates in which the HPE (Eqs.~\ref{eq:horizontalv}-\ref{eq:energyp}) are solved primarily because the vertical coordinate itself changes the meaning of vertical derivatives and the different surfaces (gas pressure vs. geometric height) on which the horizontal gradients of fields are defined.

In the planet's rotating reference frame, the momentum equation for the neutral species is given by (see \citealt{Song2001}):
\begin{eqnarray}\label{eq:neutrals}
    \rho_n\frac{d\underline{v}_n}{dt}=-\nabla p_\mathrm{gas}+\underline{F}_n+\rho_i\nu_{in}(\underline{v}_i-\underline{v}_n)+\rho_e\nu_{en}(\underline{v}_e-\underline{v}_n),
\end{eqnarray}
where $\underline{v}_n=(v_{n,r}, v_{n,\theta}, v_{n,\phi})$	is the velocity vector of the neutral species with the radial velocity $v_{n,r}$~[$\mathrm{m\,s^{-1}}$], the zonal velocity $v_{n,\phi}$~[$\mathrm{m\,s^{-1}}$] and the meridional velocity $v_{n,\theta}$~[$\mathrm{m\,s^{-1}}$], $p_\mathrm{gas}$~[Pa] the gas pressure of the neutral species, $\underline{F}_n$ the sum of external forces acting on the neutrals (e.g., gravity, Coriolis, centrifugal), $\rho_{i}$~[kgm$^{-3}$] the ion mass density, $\rho_{e}$~[kgm$^{-3}$] the electron mass density, and $\rho_{n}$~[kgm$^{-3}$] the neutral mass density, and $\nu_{in}$, $\nu_{en}$ [s$^{-1}$] the ion-neutral and electron-neutral collision frequencies, respectively. 
The last two terms on the right-hand side represent the frictional drag resulting from the relative motion between charged and neutral species. In this paper, gas species are considered only.
The momentum Eq.~\ref{eq:neutrals} without the frictional terms in height coordinates is equivalent to the horizontal and vertical momentum (Eqs.~\ref{eq:horizontalv} and \ref{eq:vericalv}) in $p$-coordinates using the hydrostatic balance assumption in the vertical direction.

The momentum equations for electrons (subscript $s=e$) and ions ($s=i$) are given by (see, e.g.,  \citealt{Song2001}) 
\begin{align}\label{eq:electrons}
n_em_e\frac{d\underline{v}_e}{dt}&=-\nabla P_e+\underline{F}_e +en_e(\underline{E}+\underline{v}_e\times\underline{B})-n_em_e\nu_{en}(\underline{v}_e-\underline{v}_n),\\\label{eq:ions}
n_em_i\frac{d\underline{v}_i}{dt}&=-\nabla P_i+\underline{F}_i +en_e(\underline{E}+\underline{v}_i\times\underline{B})-n_em_i\nu_{in}(\underline{v}_i-\underline{v}_n),
\end{align}
where $e$~[C] is the elementary charge, $\underline{F}_s$ the external forces exerting on the charged species $s$, $n_s$~[m$^{-3}$], $P_s$~[Pa], $m_s$~[kg] the number density, pressure, and mass of each charged species.
$\underline{E}$ and $\underline{B}$ are the electric and magnetic fields.  
Here, the terms with electron-ion collision frequency $\nu_{ei}$ and ion-electron collision frequency $\nu_{ie}$ were neglected, since our estimates have shown that for our parameter regime $\nu_{en}\gg \nu_{ei}$ and $\nu_{in}\gg \nu_{ie}$.
Since \texttt{ExoRad} does not evolve ion or electron dynamics, a steady-state approximation ($\frac{d}{dt}=0$) for their drift velocities is derived using a reduced momentum balance equation for electrons and ions, assuming singly charged ions, quasi-neutrality, and neglecting pressure gradients and inertial terms (e.g., \citealt{SchunkNagy2009}):
\begin{equation} \label{ref:species_eq}   e(\underline{E}+\underline{v}_s\times \underline{B})-m_s\nu_{sn}(\underline{v}_s-\underline{v}_n)\approx 0.
\end{equation}
We further assume that the large-scale electric field in the planetary rest frame is negligible ($\underline{E}\approx 0$).
This implies that the electric field in the local frame of the neutral fluid is $\underline{E}'=\underline{E}+\underline{v}_n\times \underline{B}\approx \underline{v}_n\times \underline{B}$.
This approach does not take  the feedback from the global current system into account that would modify the large scale electric field and currents.
The main limitation of the model is that it neglects the polarization electric field and therefore does not ensure current closure.
In a partially ionized atmosphere, collisions between neutrals and charged particles introduce charge separation, which creates a polarization electric field. 
This field is  build up until it balances the differential drifts of ions and electrons, though maintaining current continuity by diverting currents within the ionosphere.
By neglecting the polarization field, the model only describes the local Lorentz force component of the coupling between charged particles and neutrals, without taking into account any self-consistent electrodynamic feedback in the ionosphere. 
Therefore, the calculated magnetic drag describes a local forcing approximation rather than the global response of an ionospheric current system. 
However, this parametrization provides a physically based and computationally efficient method to study the main large scale effects of magnetic coupling on the atmospheric dynamics and thermal structure.
The model separates the impact of direct collisional drag between charged particles and neutrals, which dominates in the upper atmosphere.
Nevertheless, a more complete approach, i.e. solving the  Poisson's equation, would enable global current closure and feedback from large-scale electric fields (see Sec. \ref{sec:discussion}).
These effects could modify the magnitude and spatial distribution of the frictional drag and heating. This, however, is outside the scope of the present paper.

We further assume a dipolar large-scale planetary magnetic field anti-aligned with the planetary rotation axis.
The field is  purely poloidal, with the magnetic flux density given by
\begin{equation}\label{eq:dipole}
    \underline{B}_\mathrm{dip}(r,\theta)=B_0\left(\frac{R_\mathrm{ref}}{r}\right)^3(-2\sin\theta\,\underline{\hat{e}}_r+\cos\theta\,\underline{\hat{e}}_\theta).
\end{equation}
To calculate the dipolar magnetic field, the equatorial magnetic field strength $B_0$~[T] is defined at the reference radius $R_\mathrm{ref}$~[m] (Sec. \ref{sec:approach}).
The northward component $B_{\mathrm{dip},\theta}$~[T] is positive in the direction of increasing $\theta$.
The magnitude of the magnetic field, $|\underline{B}_\mathrm{dip}|$,   and its unit vector, $\underline{\hat{b}}=(\hat{b}_r,\hat{b}_\theta,\hat{b}_\phi)$,  are
\begin{eqnarray} \label{eq:Bdip}
|\underline{B}_\mathrm{dip}|&=&B_0\left(\frac{R_\mathrm{ref}}{r}\right)^3\sqrt{1+3\sin^2\theta},\\
\label{eq:bhat} \underline{\hat{b}}&=&\frac{\underline{B}_\mathrm{dip}}{|\underline{B}_\mathrm{dip}|}=\frac{(-2\sin\theta\,\underline{\hat{e}}_r-\cos\theta\,\underline{\hat{e}}_\theta)}{\sqrt{1+3\sin^2\theta}}.
\end{eqnarray}
To simplify the implementation, the $r$-dependence of the magnetic field is neglected in the simulations of \texttt{ExoRad}.
Treating $\underline{B}_\mathrm{dip}$ as  constant with altitude introduces only a small error of less than 4~\% in $|\underline{B}_\mathrm{dip}|$ in the modeled atmospheric region of WASP-18~b, corresponding to a radial extent of $\Delta z\approx1075$~km.
This estimate is specific to the geometry and scale height of WASP-18 ~b. In (U)HJs with more extended atmospheres or smaller radii, the variation of $\underline{B}_\mathrm{dip}$ with altitude could be more pronounced and should be accounted for in the analysis.
Additionally, it is assumed for the drag calculation that the neutral wind has no radial component ($v_{n,r}\approx 0$~m/s), as vertical winds are  weak compared to horizontal winds.
Solving Eq.~\ref{ref:species_eq} for the drift velocity $\Delta\underline{v}_s=\underline{v}_s-\underline{v}_n$ yields
\begin{equation}\label{ref:vel_s}
    \Delta\underline{v}_s=\pm \frac{k_s}{1+k_s^2}(\underline{v}_n\times\underline{\hat{b}})-\frac{k_s^2}{1+k_s^2}\underline{v}_{n,\perp},
\end{equation}
where the upper sign applies to ions ($s=i$) and the lower to electrons ($s=e$). 
The velocity of the neutrals perpendicular to $\underline{B}_\mathrm{dip}$ is given by
\begin{equation}
    \underline{v}_{n,\perp}=\underline{v}_{n}-(\underline{v}_{n}\cdot \underline{\hat{b}})\,\underline{\hat{b}}.
\end{equation} 
Equation~\ref{ref:vel_s} shows that 
if $k_s\ll 1$, collisions dominate and the charged species are strongly coupled to the neutrals. 
In this regime, magnetic drag is weak. 
For $k_s\gg 1$, the charged particles complete many gyro-orbits before being deflected by collisions.
They are magnetized and coupled primarily to the magnetic field, leading to significant relative drift and enhanced magnetic drag on the neutrals.
Substituting Eq.~\ref{ref:vel_s} into the momentum exchange terms of Eq.~\ref{eq:neutrals}, the total drag force density exerted by charged particles on the neutral fluid is
\begin{align}
    \underline{F}_\mathrm{drag}=&\rho_i\nu_{in}\left(\frac{k_i}{1+k_i^2}(\underline{v}_n\times\underline{\hat{b}})-\frac{k_i^2}{1+k_i^2}\underline{v}_{n,\perp}\right)\nonumber \\
    &+ \rho_e\nu_{en}\left(-\frac{k_e}{1+k_e^2}(\underline{v}_n\times\underline{\hat{b}})-\frac{k_e^2}{1+k_e^2}\underline{v}_{n,\perp}\right).
\end{align}
To estimate the contributions from ion and electron drag, the Pedersen $K_{P,s}$~[$\mathrm{kg\,m^{-3}s^{-1}}$] and Hall drag coefficients $K_{H,s}$~[$\mathrm{kg\,m^{-3}s^{-1}}$] for each species are introduced:
\begin{eqnarray}\label{eq:KPs}
    K_{P,s}&=\rho_s\nu_{sn}\frac{k_s^2}{1+k_s^2},\\\label{eq:KHs}
    K_{H,s}&=\rho_s\nu_{sn}\frac{k_s}{1+k_s^2},
\end{eqnarray}
so that the total drag force density simplifies to
\begin{equation}\label{eq:drag_final}
    \underline{F}_\mathrm{drag}=\left(K_{H,i}-K_{H,e}\right)\left(\underline{v}_n\times\underline{\hat{b}}\right)-\left(K_{P,i}+K_{P,e}\right)\underline{v}_{n,\perp}.
\end{equation}
The first term represents the Hall drag, which acts perpendicular to both 
$\underline{v}_n$ and $\underline{B}_\mathrm{dip}$, while the second term corresponds to the Pedersen drag, which opposes the perpendicular motion of the neutrals to $\underline{B}_\mathrm{dip}$.
Although ion and electron velocities are not evolved dynamically, their influence is included through the Hall and Pedersen coefficients, which depend on local $x_e$,  plasma parameters, and magnetic field strength of the dipolar field.
With Eq.~\ref{eq:xe} the electron and ion mass density ($\rho_s=m_sx_en_n$) can be calculated.

\begin{figure}
\noindent\includegraphics[width=0.4\textwidth]{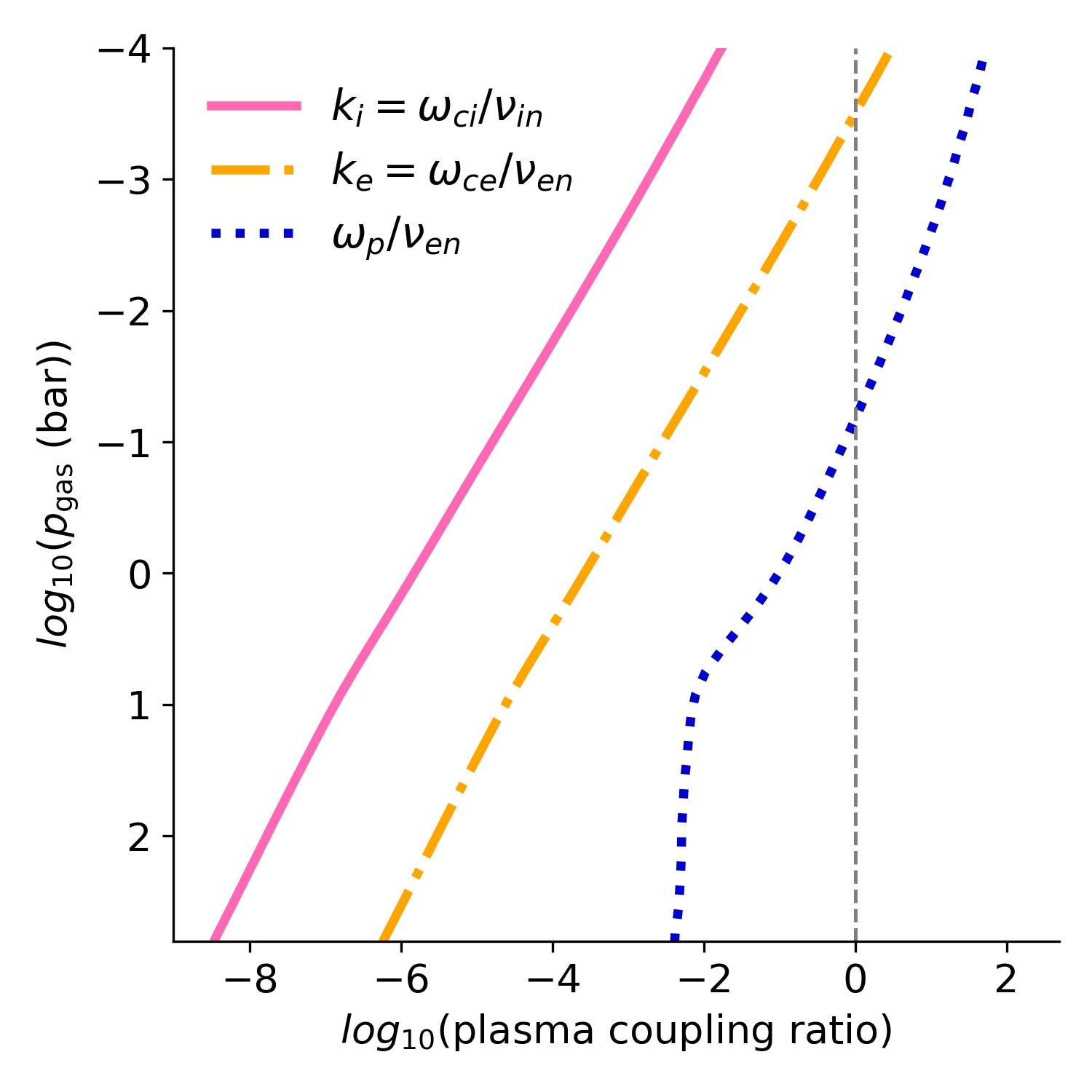}
\caption{Magnetization parameter (Eq.~\ref{eq:magnetization}) profiles for ions ($k_i$) and electrons ($k_e$) averaged over all latitudes and over dayside atmosphere ($-90$°$<\phi\leq90$°).
The blue dotted line ($\omega_{pe}/\nu_{en}$) shows the relation between   the electron plasma frequency and the electron–neutral collision frequency averaged over  the dayside atmosphere. 
The plasma coupling ratios are calculated from the \texttt{ExoRad} simulation for WASP-18 b with anisotropic drag.
 The dashed grey line shows where the plasma coupling ratio reaches unity.}
\label{Fig:k_parameter}
\end{figure}

\texttt{ExoRad} assumes vertical hydrostatic equilibrium, employing gas pressure as the vertical coordinate. It does not explicitly solve the vertical momentum equation. Instead, vertical velocities are derived to ensure mass continuity. While vertical magnetic drag forces can occur (e.g., \citealt{Christie2025}), their inclusion would violate the vertical hydrostatic equilibrium.
However, the resulting vertical acceleration is expected to be negligibly small compared to gravity in the weakly ionized regime considered here.
Therefore, the radial drag is neglected and only the meridional and zonal components of the drag force density are computed:
\begin{eqnarray} \label{eq:drag_comp}
    F_\mathrm{drag,\theta}&=&K_{H}v_{n,\phi}\hat{b}_r-K_{P}v_{n,\theta}\hat{b}_r^2\\\label{eq:drag_comp2}
    F_\mathrm{drag,\phi}&=&-K_{H}v_{n,\theta}\hat{b}_r-K_{P} v_{n,\phi},
\end{eqnarray}
where 
\begin{eqnarray}\label{eq:totalKH}
    K_{H}&=&K_{H,i}-K_{H,e}\,[\mathrm{kg\,m^{-3}s^{-1}}],\\
    \label{eq:totalKP} K_{P}&=&K_{P,e}+K_{P,i}\,[\mathrm{kg\,m^{-3}s^{-1}}] 
\end{eqnarray}
are the total Hall and Pedersen coefficients, respectively.
The geometric relation $\hat{b}_\theta^2=1-\hat{b}_r^2$ was used assuming the azimuthal component of the magnetic field is negligible ($\hat{b}_\phi=0$).
The expression in Eqs.~\ref{eq:drag_comp} and \ref{eq:drag_comp2} is implemented in \texttt{ExoRad} as an external drag force acting on the horizontal momentum Eq.~\ref{eq:horizontalv} via $\underline{F}_v=\frac{1}{\rho_n}\underline{F}_\mathrm{drag}$.
$\nu_{in}$ and $\nu_{en}$ are
\begin{eqnarray}\label{eq:nuin}
\nu_{in}&=& 4.6\times10^{-16} n_n,\\\label{eq:nuen}
    \nu_{en}&=& 6.967\times 10^{-14}n_n T_\mathrm{gas}^{0.1},
\end{eqnarray}
where $n_n$~[m$^{-3}$]  is the neutral number density.
The electron-H collision rate  follows \citet{Koskinen2010}, based on \citet{Danby1996}. 
The chemical atmosphere gas composition strongly varies between the day- and nightside. While the dayside is dominated by atomic H, the nightside is dominated by H$_2$ (Fig. 8 of \citealt{2019A&A...626A.133H}). In this work, $\nu_{en}$ is computed assuming atomic hydrogen as the dominant neutral species, which is justified on the dayside of WASP-18~b. On the nightside, where molecular hydrogen dominates, $\nu_{en}$ might be underestimated, as collision frequencies differ (Eq.~12 in \citealt{Koskinen2010}). However, the electron density and therefore the magnetic coupling strongly decrease on the nightside and the resulting impact on the magnetic drag is small.
The ion-neutral rate  is based on a rigid-sphere model for non-resonant collisions \citep{Chapman1956, Gunzkofer2023}, assuming a mean atomic mass number of 32 that was estimated by summing the relevant atomic masses that make up an ion as given in \texttt{GGchem}. 
However, the effective ion mass may vary with altitude and longitude as the dominant ion species is determined by the gas temperature and elemental abundances.
In the plasma regime considered here, ions are collisionally coupled to the neutrals ($k_i\ll1$, Fig.~\ref{Fig:k_parameter}), so the resulting variations in ion mass have a comparatively weak effect on the conductivities relative to the dominant dependence on electron density. 
Therefore, the assumed constant mass number is sufficient for the present work, however future studies including altitude-dependent ion composition would be valuable.
Resonant collisions (e.g., Na$^+$–Na) are negligible due to the dominance of hydrogen in the neutral atmosphere.

\begin{figure}
\noindent\includegraphics[width=0.4\textwidth]{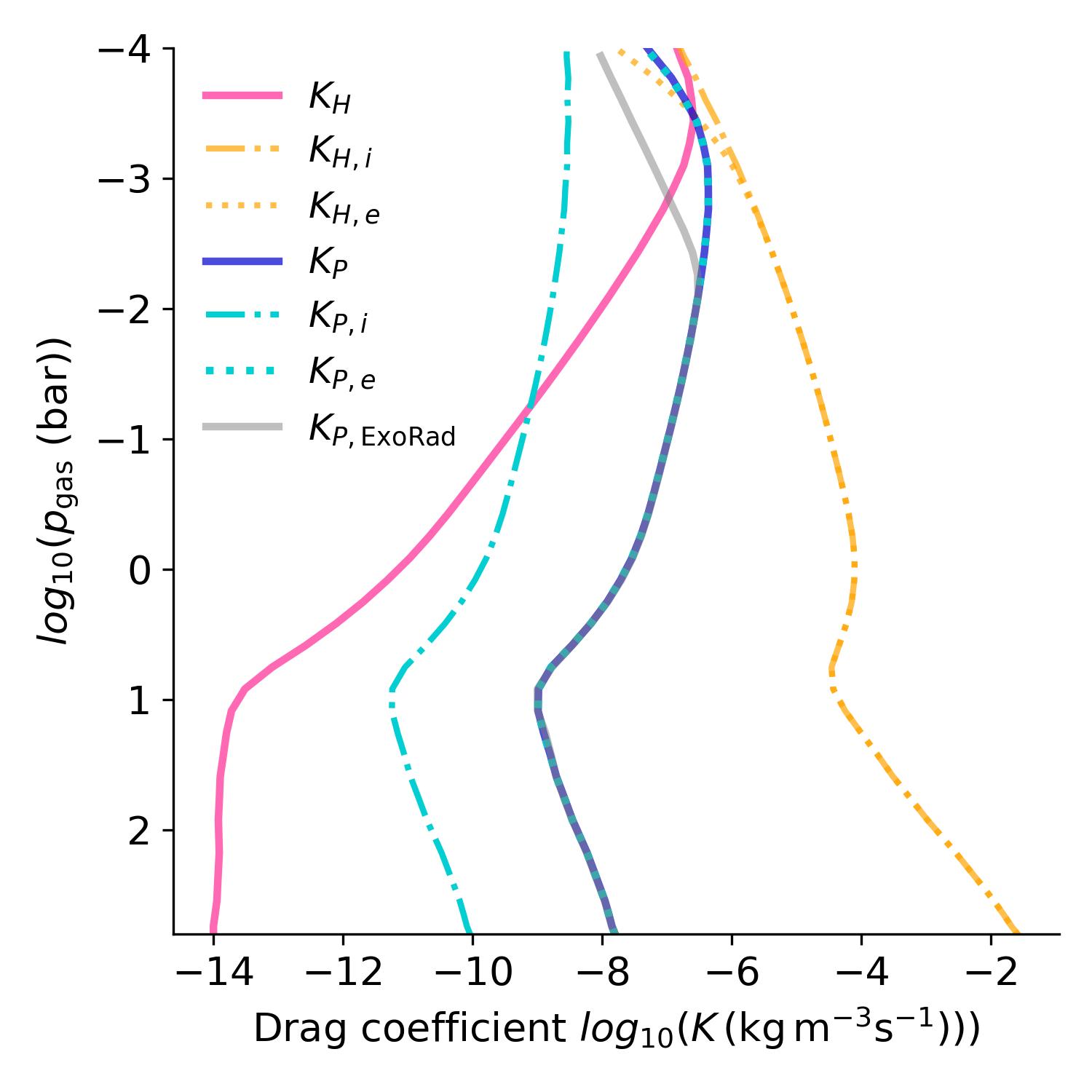}
\caption{Hall ($K_{H,s}$) and Pedersen ($K_{P,s}$) drag coefficients for electrons ($s=e$) and ions ($s=i$) 
averaged over all latitudes and over dayside atmosphere ($-90$°$<\phi\leq90$°). $K_{H}=K_{H,i}-K_{H,e}$ and $K_{P}=K_{P,i}+K_{P,e}$ are the total Hall and Pedersen coefficients.
The coefficients were calculated using the gas pressure-temperature profile from the \texttt{ExoRad} simulation run of WASP-18~b with anisotropic drag. The grey line shows $K_P$ obtained directly from \texttt{ExoRad}. It diverges from the calculated $K_P$ (blue line) in the upper atmosphere because the drag timescale in \texttt{ExoRad} is limited to $\tau_\mathrm{drag}\geq100$~s. The same applies for $K_H$ from \texttt{ExoRad} (not shown here).}
\label{Fig:drag_coeff}
\end{figure}

The  total drag force density (Eq.~\ref{eq:drag_final}) provides both a sink of momentum and a source of frictional heating, and is particularly important where the magnetization parameters $k_i$ and $k_e$ are large.
In Fig. \ref{Fig:drag_coeff}, the Hall and Pedersen drag coefficients according to Eqs.~\ref{eq:KPs} and \ref{eq:KHs}, are calculated using the gas pressure ($p_\mathrm{gas}$) and temperature ($T_\mathrm{gas}$) output from the \texttt{ExoRad} simulation of the WASP-18 b atmosphere with anisotropic magnetic drag, as well as $x_e$ output from \texttt{GGchem}. The coefficients shown are averaged over all latitudes and over the dayside atmosphere ($-90$°$<\phi\leq90$°).
For $p_\mathrm{gas}\geq 10^{-3}$~bar, $K_{H,e}$ and $K_{H,i}$ (orange lines) are of similar magnitude and largely cancel due to opposite signs ($K_H$, magenta line), making the electron Pedersen drag ($K_{P,e}$, dotted cyan line) the dominant contributor. 
At higher altitudes where $p_\mathrm{gas}<10^{-3}$~bar, $k_e>1$ and $k_i<1$ (see Fig. \ref{Fig:k_parameter}) but the small electron mass suppresses $K_{H,e}$, so $K_{H,i}$ dominates the total Hall drag ($K_{H}$) and the Hall drag dominates the total Pedersen drag ($K_P$).
To mantain the numerical stability in \texttt{ExoRad} (Sec. \ref{sec:approach}), the drag timescale is constrained to $\tau_\mathrm{drag}=\min(\rho_n/K_H\,, \rho_n/K_P)\geq100$~s.
Consequently, the imposed lower limit reduces the effective drag strength at low pressures.
Therefore, the dayside-averaged $K_P$  deviates from its physically expected value at  $p_\mathrm{gas}<10^{-2}$~bar  (gray line in Fig. \ref{Fig:drag_coeff}), and $K_H$ at $p_\mathrm{gas}\lesssim10^{-3}$~bar.

In addition to affecting neutral momentum, magnetic drag contributes to energy dissipation via frictional heating.
The heating of the neutral atmosphere due to magnetic drag is often referred to as Joule or Ohmic heating, but it is more accurately described as frictional heating resulting from momentum exchange between drifting charged particles and neutrals and not electromagnetic dissipation, which results from electric currents interacting with the electrical resistivity of the gas \citep{VasyliunasSong2005}.
For energy to be conserved the corresponding rate at which energy is dissipated into heat per unit mass of the neutral gas is equal to the rate at which Pedersen drag force works on the fluid, given by
\begin{equation}\label{eq:qfric}
    q_\mathrm{fric}=-\frac{1}{\rho_n}\underline{F}_\mathrm{drag}\cdot \underline{v}_{n}=\frac{1}{\rho_n} K_P\underline{v}_{n,\perp}^2.
\end{equation}
$q_\mathrm{fric}$ is always positive, as it represents a conversion of neutral kinetic energy into thermal energy.
The Hall drag force is a deflective force, it changes the direction of the flow but does not remove kinetic energy from the flow.
The Pedersen drag component is frictional and removes kinetic energy from the flow which converts into heat.
To include the frictional heating in the energy equation in \texttt{ExoRad},
$q_\mathrm{fric}$ is evaluated at each time step and added to the potential temperature tendency equation. 
\section{Impact of magnetic drag on winds and temperatures in the atmosphere of WASP-18~b}\label{sec:results}
\begin{figure*}
\noindent\includegraphics[width=1\textwidth]{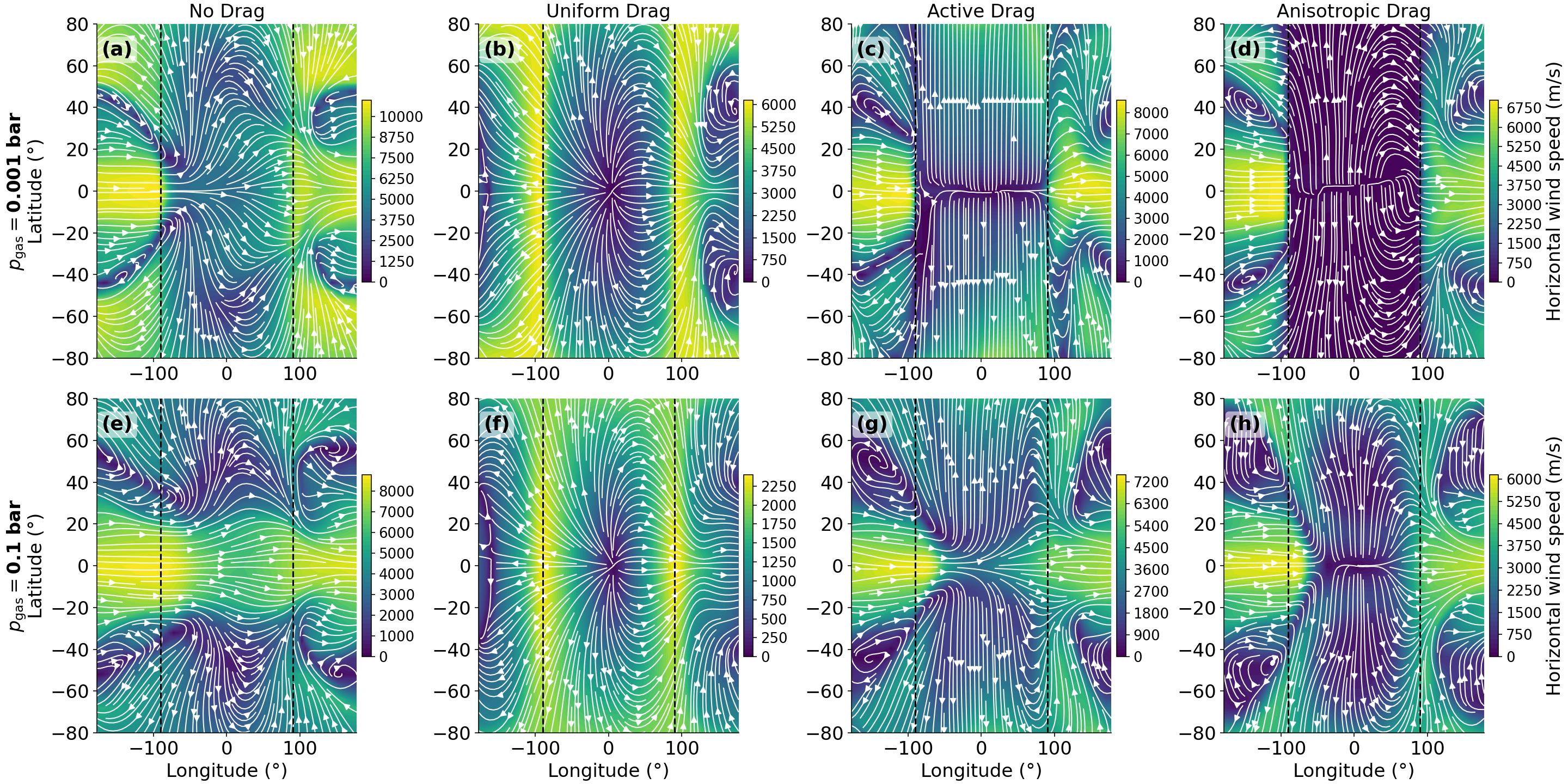}
\caption{Maps of horizontal wind speed ($|\underline{v}_h|$) and streamlines (white lines) at two pressure levels ($p_\mathrm{gas}=$0.1~bar and $p_\mathrm{gas}=$0.001~bar) for the four different magnetic drag treatments (no drag, uniform drag, active drag, and anisotropic drag). 
The substellar point is centered at each map.
The dashed black vertical lines mark the area at longitude $\pm 90^\circ$.
The output of each velocity component is time-averaged over 100 days of simulation time to eliminate the small-scale fluctuations.} 
\label{Fig:wind_speed}
\end{figure*}
\begin{figure*}
\noindent\includegraphics[width=1\textwidth]{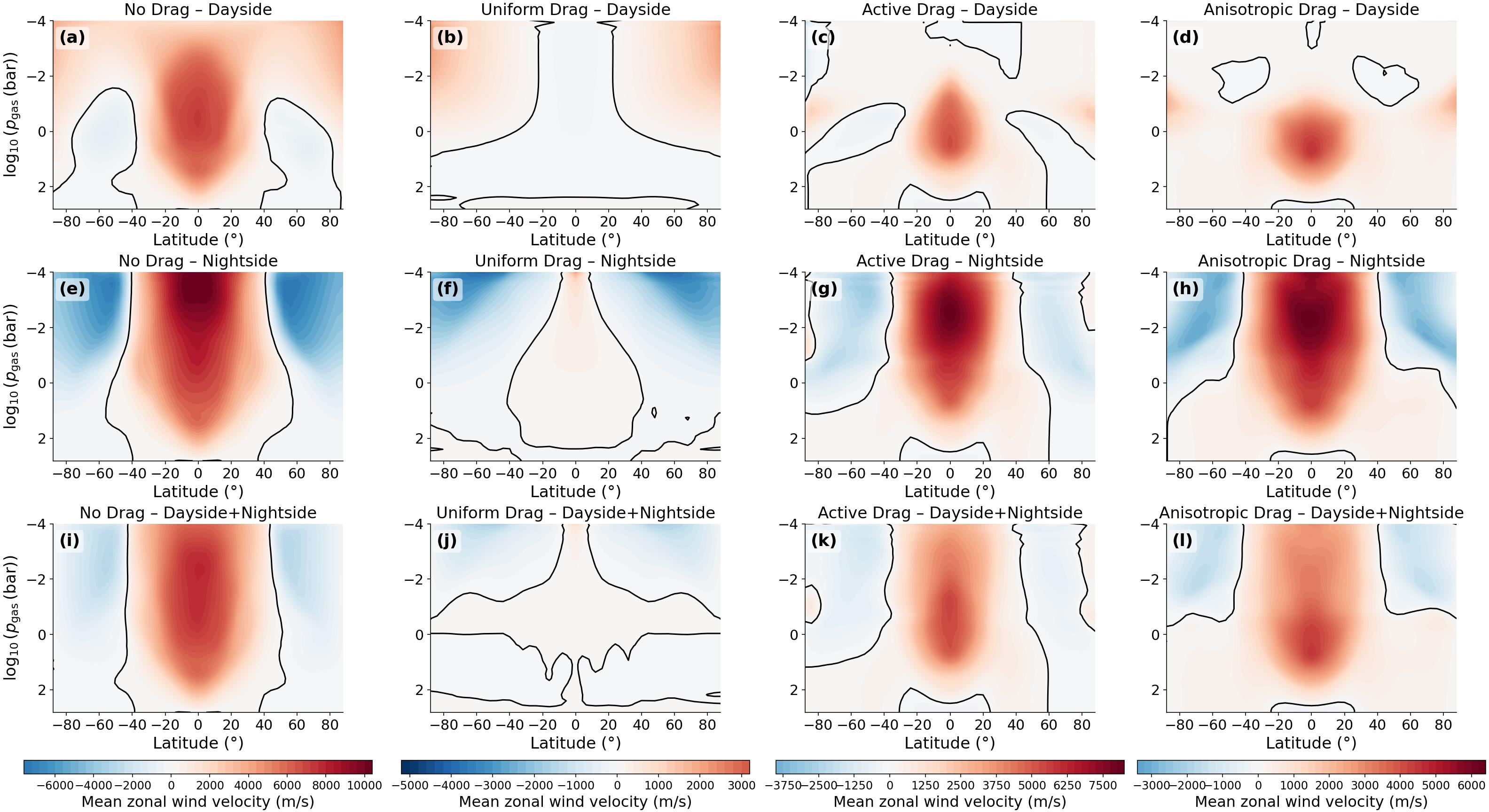}
\caption{Mean zonal wind for the four different magnetic drag treatments in different regions of the atmosphere: dayside atmosphere (first row), nightside atmosphere (second row), and total atmosphere (third row). 
The zonal wind velocity $u$ is averaged at a specific latitude, and pressure level across longitudes on the dayside ($-90$°$\leq\phi\leq 90$°), on the nightside ($|\phi|>90$°), and all longitudes.
The black line shows the zero-wind contour, the boundary between superrotation and counterrotation. 
The output of the zonal wind velocity is time 
averaged over 100 days of simulation time to eliminate the small-scale fluctuations.} 
\label{Fig:zonal_u}
\end{figure*}
\begin{figure*}
\noindent\includegraphics[width=1\textwidth]{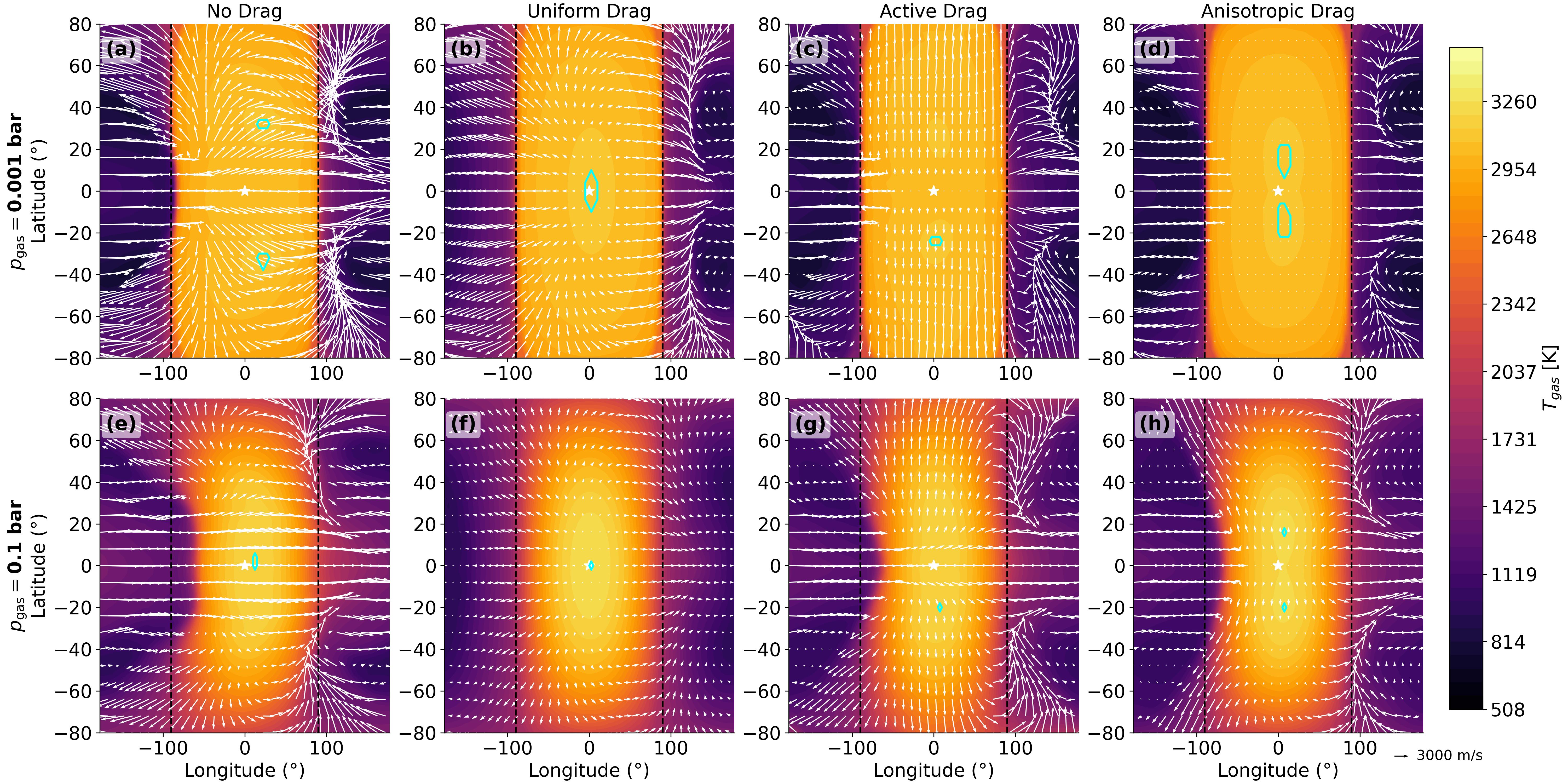}
\caption{Maps of gas temperature and velocity field at two gas pressure levels ($p_\mathrm{gas}=0.1$~bar and $p_\mathrm{gas}=0.001$~bar) for the four different magnetic drag treatments. 
The substellar point is centered at each map and is marked with a white star.
The velocity field is shown by the white arrows. The length of the arrow scales with the relative wind velocity.
The dashed black vertical lines mark the area at longitude $\pm 90^\circ$.
The cyan contours represent the hotspot region ($T_\mathrm{gas,max},T_\mathrm{gas,max}-1$K) with $T_\mathrm{gas,max}$ being the largest temperature for the chosen pressure level. The temperature is time averaged over 100 days of simulation time to eliminate the small-scale fluctuations.} 
\label{Fig:hotspot}
\end{figure*}
\begin{figure}
\noindent\includegraphics[width=0.5\textwidth]{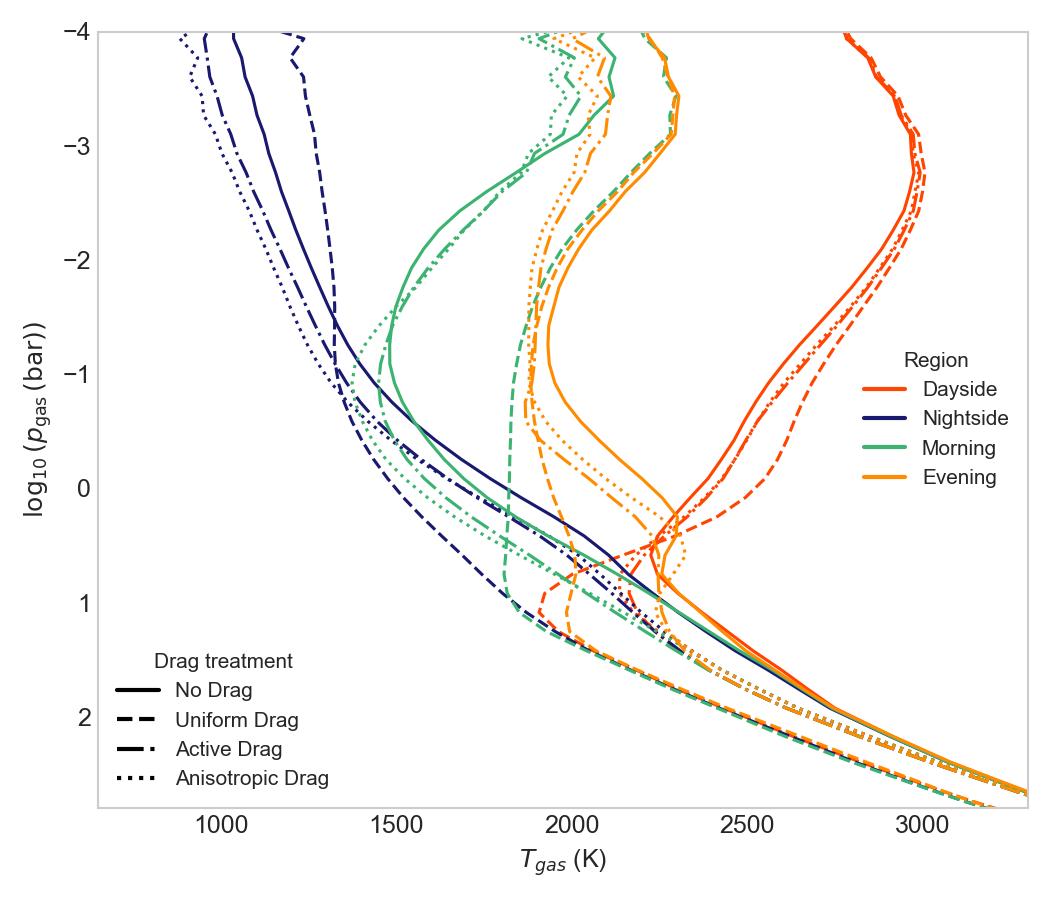}
\caption{Gas Temperature-Pressure profiles for four different drag treatments averaged over all latitudes and over different regions in the atmosphere: dayside ($-90$°$<\phi\leq90$°), nightside ($|\phi|>90$°), morning terminator ($-97.5$°$\leq\phi\leq-82.5$°) and evening terminator ($82.5$°$\geq\phi\geq97.5$°). All drag treatments show pronounced day-night differences and temperature inversions in the upper atmosphere. Nightside temperatures are cooler when active or anisotropic magnetic drag is applied compared to no drag.} 
\label{Fig:temp_profile}
\end{figure}
Magnetic drag is expected to be crucial in shaping UHJs atmospheres, yet its effects remain insufficiently studied and constrained.
Here, it is shown how magnetic drag modifies the wind pattern and thermal structure in the atmosphere of WASP-18 b and how different magnetic drag models compare to each other.
It will be explored to what extent the magnetic drag disrupts the equatorial superrotation, how it modifies heat redistribution between day and night, and whether it can change the flow pattern.
The results show that magnetic drag generally weakens equatorial superrotation, and that anisotropic drag can lead to a flow decoupling and deflection between the day- and nightside.
\subsection{Atmospheric circulation and different drag treatments}
Figure~\ref{Fig:wind_speed} shows the horizontal wind speeds and streamlines for the different drag treatments ('no drag', 'uniform drag', 'active drag', and 'anisotropic drag') for WASP-18~b as test case planet. 
Figure~\ref{Fig:zonal_u} shows the corresponding mean zonal wind structures.
In the absence of magnetic drag (when momentum and energy transfers are unaffected by the magnetic drag), the circulation is dominated by a pronounced eastward (superrotating) equatorial jet extending across the planet.
However, the strength of the jet is weakened in the upper dayside atmosphere ($p_\mathrm{gas}\leq10^{-3}$ bar: Fig. \ref{Fig:wind_speed}a, \ref{Fig:zonal_u}a) likely as a result of radiative cooling and heating or vertical wind shear.
The upper atmosphere is more strongly affected by direct stellar irradiation and less by advection compared to the denser lower atmosphere.
The strong irradiation heats then the lower atmosphere and determines the thermal structure via radiative processes. 
Steep thermal gradients from irradiation create vertical wind shear of the zonal wind. 
Such shear leads to momentum redistribution and weakening of the zonal jet in the upper atmosphere. 
Thus, radiative processes play an important role in the upper atmosphere, while advective heat redistribution is more dominant in the lower atmosphere.
This circulation pattern is common in hot Jupiter simulations without magnetic drag (e.g., \citealt{Showman2009,Showman2010,ShowmanPolvani2011}), driven by a Matsuno–Gill–type response \citep{Matsuno1966,Gill1980}.
Peak horizontal wind speeds exceed 10 km/s.

If a constant magnetic drag acts uniformly on the horizontal momentum and energy transfer ('uniform drag'), the equatorial superrotating jet is largely suppressed (Figs. \ref{Fig:wind_speed}b, \ref{Fig:wind_speed}f, \ref{Fig:zonal_u}j). 
The circulation is instead dominated by eastward and westward flows of comparable magnitude, which mostly cancel in the zonal mean (Fig. \ref{Fig:zonal_u}j).
Similar to \cite{Beltz2022} (their figure 8 for the total atmosphere), a weaker eastward jet is present at $p_\mathrm{gas}<1$ bar in higher latitudes ($|\phi|>50$°) on the dayside atmosphere (Fig. \ref{Fig:zonal_u}b).
On the nightside (Fig. \ref{Fig:zonal_u}f), a weak and narrow equatorial jet is visible for $p_\mathrm{gas}<10^{-3}$ bar.
The circulation is concentrated near the day–night terminators (longitudes $\approx \pm 90$°), with enhanced horizontal wind speeds. 
The flow diverges out of the substellar region.
Uniform drag removes momentum throughout the atmosphere uniformly, damping the wave–driven feedback which sustains  the equatorial superrotation. 
Hence, a strong global uniform drag damps the equatorial jet throughout the entire atmosphere. 
However, such a model does not capture the underlying physics, since it neglects the strong spatial variability of the ionization fraction particularly on the day- and nightside resulting in magnetic coupling of both, the dayside and nightside.

A drag confined only to the zonal direction  ('active drag': Figs.~\ref{Fig:wind_speed}c, \ref{Fig:wind_speed}g) opposes the zonal flow, decouples  the day- to nightside flow completely (at $p_\mathrm{gas}<10^{-3}$~bar), and therefore suppresses the equatorial jet on the dayside (Fig.~\ref{Fig:zonal_u}c).
At low pressures higher in the atmosphere, where drag timescales are extremely short, the zonal (east–west) flow is efficiently damped and the flow on the dayside is redirected toward the poles, producing a largely meridional (north–south) circulation on the dayside (Figs.~\ref{Fig:wind_speed}c, \ref{Fig:zonal_u}c), consistent with previous studies (e.g., \citealt{Beltz2022}).
A zonal magnetic drag is based on the assumption that only meridional currents affect the atmospheric dynamics.
The vanishing of the zonal magnetic drag at the equator due to the 
$\sin\theta$ dependence in Eqs.~\ref{eq:dragt} and \ref{eq:dragtPe} has previously been discussed by \cite{Christie2025} and is visible in the streamline in the zonal direction at $\theta=0$° on the dayside in the upper atmosphere (Fig.~\ref{Fig:wind_speed}c. 
At higher pressures ($p_\mathrm{gas}=10^{-1}$ bar, Fig.~\ref{Fig:wind_speed}g, where drag is weaker, zonal flow partially recovers, but meridional flow remains the dominant component.

A more complete representation of the effect of a planetary dipole magnetic field is achieved with a magnetic drag force acting in both, the zonal and meridional directions ('anisotropic drag').
The drag strongly damps the flow especially at low gas pressures ($p_\mathrm{gas}=10^{-3}$ bar) on the dayside (Figs.~\ref{Fig:wind_speed}d,  \ref{Fig:zonal_u}d). 
The resulting magnetic drag forces effectively isolate the dayside from the nightside circulation at the equatorial region at the morning terminator ($\phi\approx-90$°, Fig. \ref{Fig:wind_speed}d).
At higher latitudes ($\phi>|30^\circ|$) at the morning terminator,
the gas flow is restored from the dayside to the nightside.
At the evening terminator ($\phi\approx90$°), the streamlines are directed from the dayside to the nightside. 
Despite this strong damping, an eastward equatorial jet remains on the nightside and extends over a broader latitude range than in the active drag case (Fig.~\ref{Fig:zonal_u}h).
The wind speed shows dayside stagnation with limited redistribution, coupled to a nightside jet sustained by momentum convergence from deeper levels. 
This indicates that anisotropic drag firstly acts to dissipate kinetic energy, slow down and deflect the flow on the dayside, while still allowing the equatorial jet to be maintained on the nightside.
The Pedersen drag primarily suppresses zonal winds, producing significantly reduced flow speeds on the dayside and the Hall drag generates a non-dissipative magnetic coupling between the meridional and zonal flow components (Eqs.~\ref{eq:drag_comp} and \ref{eq:drag_comp2}). 
In the upper atmosphere, where the Hall drag strength becomes comparable to the Pedersen drag, the Hall drag introduces an asymmetry in the flow pattern. This asymmetry is visible in Fig.~\ref{Fig:wind_speed}d in the dayside streamlines between 
$-90^\circ<\phi<0^\circ$ and $0^\circ<\phi<90^\circ$. 
In the longitude range $-90^\circ< \phi<0^\circ$, the Hall drag deflects the flow toward more polar-directed motion, whereas in $0^\circ<\phi<90^\circ$ the streamlines show increased curvature and are redirected toward the evening terminator. 
If the magnetic dipole is aligned with the planetary rotation axis (not shown here), $\hat{b}_r$  (Eq.~\ref{eq:bhat}) changes sign while maintaining the same magnitude. 
In that case, the Pedersen damping does not change, but the magnetic coupling produced by the Hall drag between $v_{n,\theta}$ and $v_{n,\phi}$  reverses sign, leading to a circulation pattern, which is mirrored on the dayside with enhanced curvature occurring in 
$-90^\circ <\phi<0^\circ$ 
and more meridional flow in 
$0^\circ<\phi<90^\circ$. 
This behavior demonstrates the strong dependence of the atmospheric flow on the magnetic field geometry in the upper atmosphere.
The resulting flow asymmetry determines the distribution of momentum across the terminator regions and indirectly affects the flow on the nightside. The flow perturbations due to the Hall effect on the nightside do not arise from strong local Hall drag, which is weak due to low ionization fraction, but are instead a non-local response from the redirected momentum on the dayside.
At the poles on the dayside where the magnetic field lines are nearly vertical the flow experiences a stronger drag in the meridional direction than at the equator where the field lines are horizontal,  meridional magnetic drag force vanishes, and only the zonal velocity is damped by the Pedersen drag component.
This is seen in Eqs.~\ref{eq:drag_comp} and \ref{eq:drag_comp2}, where the polar flow structure is produced by the dominance of the radial magnetic field component, which reaches $|\hat{b}_r| =2$ at the poles.
Both Pedersen (in the meridional direction of the magnetic drag force) and Hall drag terms are increased in the polar regions.
The Pedersen drag provides strong damping, particularly in the meridional direction, where the damping scales as $\hat{b}_r^2$, producing strong suppression of polar motions. On the other hand, the Hall drag couples the zonal and meridional velocity components through terms proportional to  $\hat{b}_r$, creating a rotation of the horizontal momentum.  The Hall term rotates the velocity vector against strong Pedersen damping, producing the flow curvature and directional asymmetry seen in the polar regions (Fig.~\ref{Fig:wind_speed}d).
At gas pressures of 1 mbar, the overall  wind structure is similar with that shown by \cite{Christie2025} (their Fig.~13). However, unlike in the results shown here, the wind flow in \cite{Christie2025} for their anisotropic approach does not experience longitudinal asymmetries due to the Hall drag and the partial decoupling of the equatorial jet at the morning terminator. The hottest planet considered in \cite{Christie2025} reaches lower dayside temperatures than WASP-18~b, resulting in a negligible contribution of the Hall term in the drag force.

In both cases, 'active drag' and 'anisotropic drag', the equatorial jet reestablishes on the dayside for $p_\mathrm{gas}>10^{-2}$ bar ('active drag', Fig. \ref{Fig:zonal_u}c) and $p_\mathrm{gas}>10^{-1}$ bar ('anisotropic drag', Fig. \ref{Fig:zonal_u}d). 
While the variation of jet width with pressure is similar in the 'anisotropic drag' and 'no drag' cases (compare Fig.~\ref{Fig:zonal_u}e and \ref{Fig:zonal_u}h), the jet in the 'active drag' case is narrower. 
\cite{Plaschzug2025} introduced the width of the equatorial jet as well as the wind speed as two of four climate characteristics to help characterizing the potential diversity of climate regimes in the ensemble of gas giant exoplanets (see their Fig. 5). Table~\ref{tab:drag_summary} shows how jet width and wind speeds are affected by magnetic drag.
The comparison between the parameters of a planet with F host star and a global temperature of $~$2400~K from Fig.~5 of \cite{Plaschzug2025} and the calculated parameters from this study for the case when magnetic effects are absent ('no drag', Tab.~\ref{tab:drag_summary}), shows that the wind jet width is significantly larger, the evening-morning temperature difference is smaller, and the day-nightside temperature difference is larger in this study.
This discrepancy might come from different initial conditions for the planet and the exclusion of TiO and VO in the  simulations for UHJs in \cite{Plaschzug2025}.
Including TiO and VO leads to  upper dayside atmosphere thermal inversion as seen, e.g., in the red line of Fig. \ref{fig:rm_xe}.
\cite{Deline2025} have shown that including magnetic drag and the strong local gas-phase absorber TiO and VO are important to fit the data from optical to IR wavelengths for WASP-18 b. 
The wind  velocity is the highest when the magnetic drag is absent and lowest when the drag force acts uniformly on the horizontal flow
implying that the magnetic drag significantly reduces the maximum wind speeds.
\subsection{Gas temperature and different drag treatments}
In Fig. \ref{Fig:hotspot},  horizontal temperature maps for different drag treatments and pressure levels and the hotspot regions are shown.
When active drag, anisotropic drag or no drag is applied to the atmospheric flow, a hotspot splitting  in the upper atmosphere ($p_\mathrm{gas}=10^{-3}$ bar) is observed almost symmetric around the equator where two distinct temperature maxima regions instead of one appear.
This kind of splitting was also seen in other models of e.g., \citealt{Carone2020, KomacekShowman2016}.
In Fig. \ref{Fig:hotspot}e, a strong eastward equatorial jet is visible (see white arrows) which transports heat efficiently from the substellar point eastward resulting in an eastward shift of the peak temperature (see cyan contours).
The wind direction around the equator is almost purely zonal. 
When this wind flow slightly changes so that wind also points in the meridional direction, these winds shear the hottest region apart and redistribute it meridionally (Fig. \ref{Fig:hotspot}a).  
The hotspots can be seen then as two symmetric lobes around the equator.
If magnetic drag acts uniformly in the atmosphere, the hotspot area is almost centered at the substellar point because the equatorial jet is suppressed (Fig. \ref{Fig:hotspot}b and \ref{Fig:hotspot}f).
Here, the momentum damping is  constant.

The 'active' and 'anisotropic drag' introduce a nonlinear feedback which arises because neutral winds adjust the induced electric field ($\underline{E}'=\underline{v}_n\times\underline{B}$), which drives currents.
These currents depend on the locally varying conductivities. 
Then they act back on the winds via damping and Hall deflection (only in the anisotropic drag case).
Therefore, these cases suppress equatorial superrotation on the dayside at low pressures (Figs.~\ref{Fig:zonal_u}c,  \ref{Fig:zonal_u}d) and distort zonal flows leading to no efficient eastward advection of heat.
When strong magnetic drag (Pedersen + Hall) acts on the flow, it damps the wave response. The circulation then  becomes less wave-driven and more controlled by pressure-gradient forces.
In Tab. \ref{tab:drag_summary} the location of the maximum temperatures for the different drag treatments at $p_\mathrm{gas}=10^{-1}$ bar is summarized.
In the case of 'active' and 'anisotropic drag', a shift of $T_\mathrm{max}$ eastwards of the substellar point by 7.5° is seen, which is less 
compared to the case when no magnetic drag is applied to the atmosphere (12.5°).
The reason is that the magnetic drag suppresses or reduces the eastward equatorial jet.
This was also proposed by previous studies, e.g., \cite{RogersKomacek2014, Kataria2015}.
In the 'active drag' and 'anisotropic drag' an asymmetry of the hotspot regions emerges (Figs. \ref{Fig:hotspot}c, \ref{Fig:hotspot}d, \ref{Fig:hotspot}g, \ref{Fig:hotspot}h). 
In the 'active drag' model, the primary hotspot region is displaced by $\sim20$° south of the equator.
A secondary hotspot region develops $\sim20$° north of the equator, but its temperature remains about 10 K cooler than the southern hotspot at $p_\mathrm{gas}=10^{-1}$ bar. Hence, locally changing drag forces and energy inputs may introduce another level of observable asymmetries, namely in form of a primary and a secondary hotspot region. 
Furthermore, the transition region between the cooler nightside and hot dayside at the morning terminator is shifted eastward at $p_\mathrm{gas}=10^{-1}$ bar around the equatorial region for the 'no drag', 'active drag', and 'anisotropic drag' cases with the shift in the 'no drag' being the strongest and in the 'anisotropic drag' being the weakest (compare the transition region to the position of the black dashed line at $\phi=-90$° in Fig. \ref{Fig:hotspot}e, \ref{Fig:hotspot}g, and \ref{Fig:hotspot}h).
A similar eastward shift is observed at the evening terminator.
This eastward shift of the day–night boundary appears due to the zonal advection of heat by the equatorial jet, which transports the thermal hotspot downstream of the substellar point. 
When magnetic drag is introduced, Lorentz forces act against the zonal flow and dissipate kinetic energy, thereby weakening the jet and reducing the longitudinal offset of the thermal transition region. 
Consequently, stronger or more spatially extended drag ('anisotropic drag') suppresses the eastward shift of the terminator.

Fig. \ref{Fig:temp_profile} shows the gas temperature-pressure profiles averaged across all latitudes within different  longitude regions for the four drag treatments.
All models show a strong difference between the day- and nightside in the upper atmosphere, with the dayside reaching almost 3000 K and the nightside remaining below 1100 K.
The profiles of the terminator regions lie in between, with the evening terminator being warmer than the morning terminator for the 'no drag', 'active drag', and 'anisotropic drag' models, consistent with the eastward-directed flow that advects hot material.
When a spatially uniform magnetic drag is applied, the terminator profiles almost overlap for $p_\mathrm{gas}<10^{-1}$ bar.
The temperature differences between the day-/ nightside and evening/morning terminator, the two remaining climate characteristics (\citealt{Plaschzug2025}), are displayed in Tab. \ref{tab:drag_summary}.
It shows that the difference between the averaged temperatures on the morning and evening terminator at $p_\mathrm{gas}=10^{-1}$~bar is strongest in the 'anisotropic drag' with the evening terminator being 20\% hotter than the morning terminator, and weakest in the 'uniform drag' treatment.
Temperature inversions are present on the dayside and absent on the nightside.
On the nightside, the magnetic drag also affects the temperature profiles.
The temperatures are lower when magnetic drag is acting on the atmospheric dynamics ('active drag' and 'anisotropic drag') compared to the case when magnetic field does not affect the dynamics ('no drag').
The nightside temperature of the 'uniform drag' model  is larger than for the other cases and almost constant for $p_\mathrm{gas}<10^{-1}$ bar, which is due to the suppression of zonal heat redistribution.
In the 'uniform drag' case, the temperature inversion at the dayside is stronger and lies deeper in the atmosphere ($p_\mathrm{gas}=10$ bar) than for the other drag treatments.
This is also the case for the evening and morning terminator regions.
At $p_\mathrm{gas}>10$ bar, the profiles converge across the regions for all drag treatments, indicating that the deeper atmosphere is much less affected by the drag. 
\begin{table*}[htbp]
\centering
\caption{\texttt{ExoRad} results: Effect of magnetic drag on WASP-18~b's climate state characteristics.}
\label{tab:drag_summary}
\begin{tabular}
{p{4.5cm}p{2.9cm}p{2.9cm}p{2.9cm}p{2.9cm}}
\hline
\textbf{Properties} & \textbf{No Drag} & \textbf{Uniform Drag} & \textbf{Active Drag} & \textbf{Anisotropic Drag} \\
\hline
Max wind speed at $10^{-3}$ bar& 10971~m/s & 6089~m/s & 8532~m/s & 7008~m/s \\
Jet width\tablefootmark{a}  at $10^{-1}$ bar& 54° & - & 39° & 54° \\
Jet width\tablefootmark{a} at $10^{-3}$ bar& 56° & - & 52° & 56° \\
$T_\mathrm{gas,max}$ at $10^{-1}$ bar & 3201~K  & 3255~K & 3197~K  &3218~K  \\
$T_\mathrm{gas,max}$ location at $10^{-1}$ bar &  12.5° east & 2.5° east & 7.5° east, 20° south &7.5° east, \newline 16°~north/ 20°~south \\
$T_\mathrm{gas,max}$  at $10^{-3}$ bar & 3085~K  & 3090~K & 3084~K  &3088~K  \\
$T_\mathrm{gas,max}$ location at $10^{-3}$ bar &  27.5° east, \newline 32° north/ south & 2.5° east & 2.5° east, 24° south &7.5° east, \newline 16° north/ south \\
Day-night  difference\tablefootmark{b} (relative) at $10^{-1}$ bar& 1208 K (0.46) & 1393 K (0.51) & 1338 K (0.5) & 1375 K (0.52) \\
Day-night difference\tablefootmark{b} (relative) at $10^{-3}$ bar& 1832 K (0.60) & 1726 K (0.57)& 1934 K (0.65) & 1971 K (0.66) \\
Evening-morning difference\tablefootmark{c} (relative) at $10^{-1}$ bar& 455 K (0.23) & 42 K (0.02)& 438 K (0.23) & 498 K (0.27) \\
Evening-morning difference\tablefootmark{c} (relative) at $10^{-3}$ bar& 328 K (0.15) & 3 K (0.002)& 145 K (0.07) & 110 K (0.05) \\
\hline
\end{tabular}
\tablefoot{The table shows the modeling results for different climate state characteristics: maximum wind speed, equatorial jet width, day-night gas temperature difference, evening-morning gas temperature difference,  and the maximum (or hotspot) gas temperature  $T_\mathrm{gas,max}$ and its geographic location.\\
\tablefoottext{a}{Jet width is calculated with the full width at half maximum for the total atmosphere and is given in degrees of latitude.}
\tablefoottext{b}{Difference between the
day ($T_\mathrm{day}$) and nightside ($T_\mathrm{night}$) averaged temperatures. Their relative difference is calculated by $(T_\mathrm{gas,day}-T_\mathrm{gas,night})/T_\mathrm{gas,day}$.}
\tablefoottext{c}{The difference between the
evening ($T_\mathrm{gas,evening}$) and morning terminator ($T_\mathrm{gas,morning}$) averaged temperatures. Their relative difference is calculated by $(T_\mathrm{gas,evening}-T_\mathrm{gas,morning})/T_\mathrm{gas,evening}$.}}
\end{table*}
\section{Discussion}\label{sec:discussion}
The implementation of the anisotropic magnetic drag demonstrates that magnetic effects can significantly reshape the atmospheric circulation of UHJs. 
The main outcome is that the magnetic drag disrupts the equatorial superrotation on the dayside, thereby decoupling the dayside and nightside circulations at the equatorial region in the upper atmosphere, deflecting the flow on the dayside, and altering the efficiency of heat redistribution. The drag dissipates kinetic energy and dampens the flow, while maintaining an eastward jet of similar width on the nightside compared to an atmosphere which is not affected by the magnetic drag.
Additionally, the dayside–nightside temperature difference increases and the hotspot and terminator shift toward the east becomes weaker.
The implementation of anisotropic magnetic drag demonstrates an important step toward including plasma-neutral interactions in \texttt{ExoRad}, but several model limitations affect the interpretation of the simulation results.  
\paragraph{Large-scale planetary electric field:}
A main simplification in the present model is the assumption of negligible large-scale planetary electric potential  ($\underline{E}_\mathrm{pol}=0$) in the planetary rest frame. 
In reality, both dynamo-generated and polarization electric fields can arise from charge separation and currents. 
These fields influence the charged particle velocities and hence the collisional drag felt by neutrals. 
By enforcing $\underline{E}_\mathrm{pol}=0$, the charged particle's motion is controlled by $\underline{v}_n\times\underline{B}$.
This includes the direct collisional drag but excludes feedback from electric fields.
A self-consistent approach of the electric field would require solving a complex elliptic equation (Poisson's equation) for the global polarization electric field across the atmosphere at every time step (e.g., \citealt{Koskinen2014}).
Therefore, the current continuity equation ($\nabla\cdot\underline{j}=0$ with the current density $\underline{j}=\underline{\underline{\sigma}}\cdot\underline{E}'$, the conductivity tensor $\underline{\underline{\sigma}}$, and the electric field seen in the frame of the neutrals $\underline{E}'=\underline{E}_\mathrm{pol}+\underline{v}_n\times\underline{B}$) has to be solved and the equation to be calculated then becomes $\nabla\cdot(\underline{\underline{\sigma}}\cdot\nabla\Phi_\mathrm{pol})=\nabla\cdot(\underline{\underline{\sigma}}\cdot(\underline{v}_n\times\underline{B}))$, where $\underline{E}_\mathrm{pol}$, which is electrostatic, is expressed as a gradient of a scalar potential $\underline{E}_\mathrm{pol}=-\nabla\Phi_\mathrm{pol}$.
To solve the Poisson equation is computationally expensive and beyond the scope of the present work.

\begin{figure}
\noindent\includegraphics[width=0.45\textwidth]{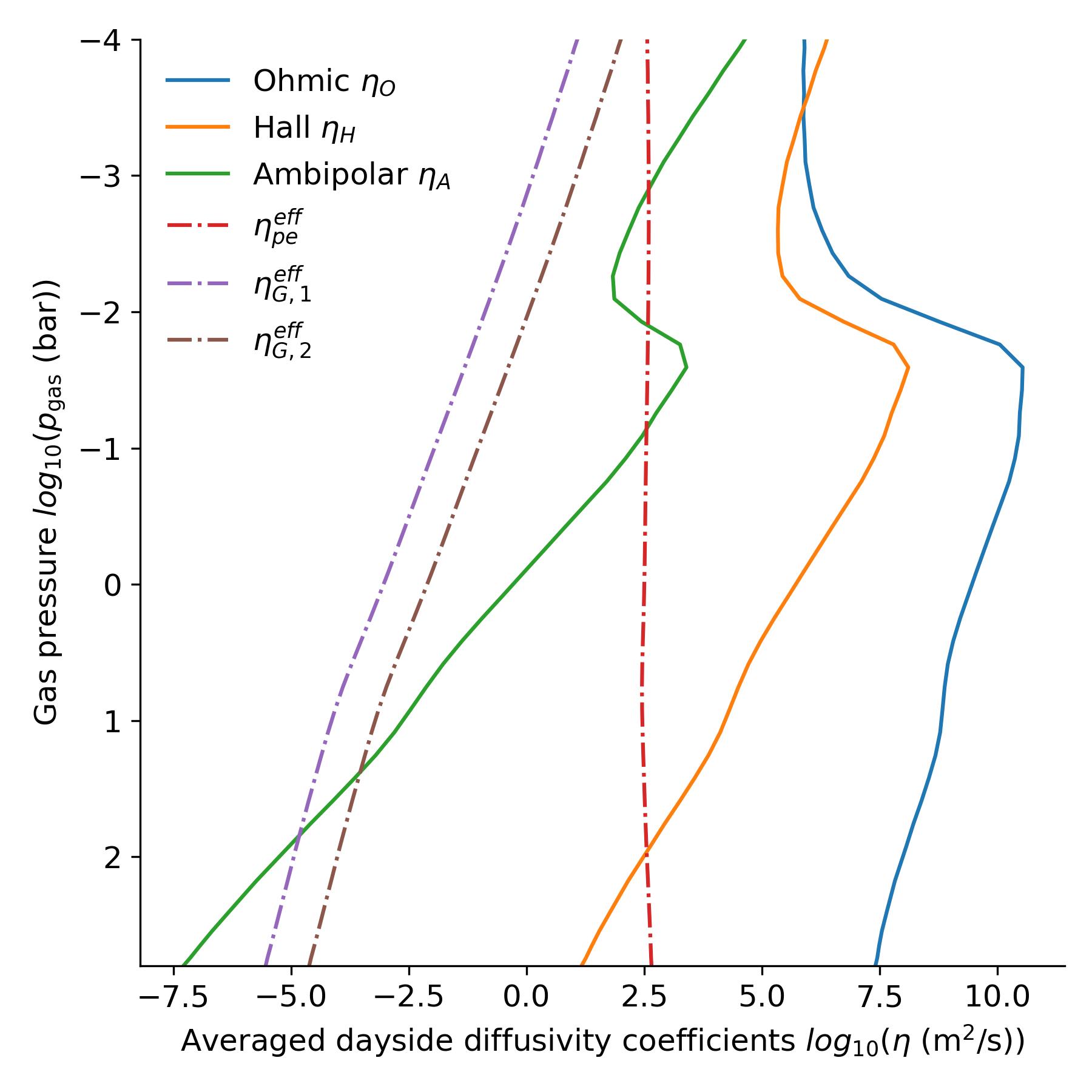}
\caption{Diffusion coefficients from the Ohm's law (Eq.~\ref{eq:ohms_law}) averaged over all latitudes and over dayside atmosphere ($-90$°$<\phi\leq90$°). The coefficients were calculated from the output of the \texttt{ExoRad} simulation with anisotropic drag for the atmosphere of WASP-18 b.}
\label{Fig:diffusion}
\end{figure}

To do a rough estimate on the error that is made by neglecting $\underline{E}_\mathrm{pol}$, the Ohm's law is considered: $\underline{j}=\sigma_P\underline{E}'_\perp+\sigma_H(\underline{\hat{b}}\times\underline{E}')+\sigma_\parallel\underline{E}'_\parallel$ where $\sigma_P$ is the Pedersen (Eq.~\ref{eq:hall_pedersen}), $\sigma_H$ is the Hall (Eq.~\ref{eq:hall_pedersen_2}), and $\sigma_\parallel$ is the parallel conductivity (Eq.~\ref{eq:el_conductivity}), and the different components of the electric field: parallel ($\underline{E}'_\parallel$) and perpendicular ($\underline{E}'_\perp$) to the magnetic field.
$\underline{E}_\mathrm{pol}$ arises from charge separation.
The initial motional electric field $\underline{E}_0=\underline{v}_n\times\underline{B}$ drives two primary currents (Pedersen: $\underline{j}_\mathrm{P,prim}=\sigma_P\underline{E}_0$, Hall: $\underline{j}_\mathrm{H,prim}=\sigma_H(\underline{\hat{b}}\times\underline{E}_0)$) and causes charges to build up at the insulating boundaries (non-conducting nightside at terminator region, non-conducting gas above and below the ionosphere). 
The charge build-up creates $\underline{E}_\mathrm{pol}$ opposing the initial charge flow.
This means that $\underline{E}_\mathrm{pol}$ drives a secondary Pedersen current ($\underline{j}_\mathrm{P,sec}=\sigma_P\underline{E}_\mathrm{pol}$) that cancels the primary Hall current ($\underline{j}_\mathrm{H,prim}=\sigma_H(\underline{\hat{b}}\times\underline{E}_0)$) to ensure $\nabla\cdot\underline{j}=0$.
A simplified calculation of the polarization electric field with approximately uniform conductivities would require to calculate $\underline{j}_\mathrm{P,sec}=-\underline{j}_\mathrm{H,prim}$ which gives $\underline{E}_p\approx-\frac{\sigma_H}{\sigma_P}\underline{\hat{b}}\times\underline{E}_0\approx -k_e\underline{\hat{b}}\times\underline{E}_0$. 
$\underline{E}_\mathrm{pol}$ drives a secondary Hall current ($\underline{j}_\mathrm{H,sec}=\sigma_H(\underline{\hat{b}}\times\underline{E}_\mathrm{pol})\approx\frac{\sigma_H^2}{\sigma_P}\underline{E}_0$). These currents flow in the same direction as the initial Pedersen drag current.
The total drag current is then $\underline{j}_\mathrm{drag,tot}=\underline{j}_\mathrm{P,prim}+\underline{j}_\mathrm{H,sec}=\sigma_P(1+\frac{\sigma_H^2}{\sigma_P^2})\underline{E}_0$.
The concept described here is analogue to the Earth's Equatorial Electojet current system (e.g.,  \citealt{Baumjohann1996, Muralikrishna2006}).
In this very simplified calculation, it was shown that neglecting $\underline{E}_\mathrm{pol}$, the anisotopic drag model underestimates both the magnetic drag force ($\underline{F}_\mathrm{drag}=\underline{j}\times\underline{B}$) and the frictional heating rate by the factor $1+k_e^2$.
$k_e$ is shown in Fig. \ref{Fig:k_parameter}. So if $k_e\approx1$ for $p_\mathrm{gas}<10^{-3}$ bar on the dayside atmosphere, then the magnetic drag force and the frictional heating are underestimated by a factor of 2. 
The model with anisotropic drag includes the initial  electric field allowing the calculation of the primary magnetic drag but it neglects the secondary fields to ensure current closure.
Therefore, the model underestimates the magnetic drag and frictional heating in the upper dayside atmosphere. 
\paragraph{Non-ideal MHD effects:}
Other magnetic effects, such as ambipolar diffusion, have also been neglected.
The non-ideal MHD effects can modify $\underline{E}$ and lead to modifications of the local Joule heating.
To evaluate their relative importance which influence the evolution of the magnetic field, the prefactors of the low-frequency Ohm’s law in the neutral frame, which is derived from the momentum equations of the electrons, ions, and the neutrals,  can be approximated by ( \citealt{Leake2014,ballester2018,Khomenko2014,forteza2007})
\begin{align}
\underline{E}^{v_n} =&
\eta_O \mu_0 \underline{j}
+ \frac{\eta_H \mu_0}{|B|}\,\underline{j}\times\underline{B}
- \frac{\eta_A \mu_0}{|B|^2}\left(\left((\underline{j}\times\underline{B}\right)\times\underline{B}\right)\nonumber \\
&- \frac{\eta_H \mu_0}{|B|}\nabla p_e
+ \frac{1}{\rho_i\nu_{in}}\Big(\nabla (2p_e) - \frac{\rho_i}{\rho_n}\nabla p_n\Big)\times\underline{B},
\end{align}\label{eq:ohms_law}
where $\underline{E}^{v_n}=\underline{E}+\underline{v}_n\times\underline{B}$ is the electric field in the neutral rest frame and $\underline{j}=\nabla\times \underline{B}/\mu_0$ is the electric current density.  
The terms on the right-hand side correspond to Ohmic diffusion, Hall effect, ambipolar diffusion, Biermann battery, and pressure-gradient contributions. 
The diffusion coefficients, scaled to units of m$^2\,$s$^{-1}$, are given by
\begin{equation}
    \eta_O=\frac{1}{\mu_0\sigma_\parallel}\approx\frac{|B|}{\mu_0e nk_e},\quad  \eta_H\approx\frac{|B|}{\mu_0e n},\quad  \eta_A\approx\frac{|B|^2}{\rho_i\nu_{in}\mu_0}.
\end{equation}
Here the assumptions $k_e \gg k_i$, $m_i \gg m_e$, $n_e \approx n_i=n$ with the plasma number density $n$~[$\mathrm{m^{-3}}$], and the plasma pressure $p=p_i+p_e\approx 2p_e$~[Pa] have been applied. 
Note that in the estimation shown here, there is no difference made between the magnetic dipole field and the induced field, so $B=B_\mathrm{dip}$. 
Effective diffusivities were introduced for the battery and pressure-gradient terms to allow comparison with the diffusive terms: $\eta^{eff}_{pe}=\frac{p_e}{|B|ne}$~[m$^2\,$s$^{-1}$], $\eta^{eff}_{G,1}=\frac{2p_e}{\rho_i\nu_{in}}$~[m$^2\,$s$^{-1}$], and $\eta^{eff}_{G,2}=\frac{p_n}{\rho_n\nu_{in}}$~[m$^2\,$s$^{-1}$]. As shown in Fig.~\ref{Fig:diffusion}, the Hall effect and Ohmic diffusion dominate in the upper dayside atmosphere ($p_\mathrm{gas}<10^{-3}$ bar), with the Hall diffusivity exceeding the Ohmic diffusivity at the upper atmosphere. The other effects remain several orders of magnitude smaller than the Hall effect and Ohmic diffusion in the region of interest.
Ambipolar diffusion has only a minor influence in the WASP-18 b case with a magnetic field strength of 5~G but possibly becomes relevant at $p_\mathrm{gas}<10^{-4}$~bar compared to Ohmic diffusion and Hall effects due to the decreasing gas number density. 
Since $\eta_A$ is dependent on $|B|^2$, increasing the magnetic field strength would enhance the effect of the ambipolar diffusion and it could become of similar magnitude than the Ohmic diffusion at, e.g. $p_\mathrm{gas}<10^{-3}$~bar for $|B|=50$~G.
However, ambipolar diffusion is linked to the feedback of the other MHD effects as well as the configuration of the dipolar and induced magnetic fields, and therefore the effect of the ambipolar diffusion might be different. 
\cite{Savel2024} proposed a new method to constrain the planetary magnetic field by measuring the ambipolar velocity in the photosphere via comparing heavy ion velocity with the neutral gas velocity using high-resolution spectroscopy. 
Therefore, this ambipolar diffusion effect can be relevant for studying 
the potential spectroscopical detectability of the ion-neutral velocity in the photosphere \citep{Soriano-Guerrero2025}.
However, the interpretation of the results of this method strongly relies on the underlying model of the drift velocity between ions and neutrals (\citealt{Christie2025}, Appendix~\ref{ap:savel}).
Pressure-gradient terms are negligible for the conditions applied here. 
\paragraph{Self-excited atmospheric dynamo:}
Another limitation of the drag parameterization is the neglect of the effect of the atmospheric dynamo. 
For $R_M\gg 1$ (Fig.~\ref{fig:RM}), atmospheric flows can produce toroidal fields comparable in strength to the imposed dipole field \citep{RogersKomacek2014}. Several possible scenarios for atmospheric magnetic induction were explored by \cite{Dietrich2022}, indicating that both the strength and geometry of the Lorentz force could show large variation from a static dipole configuration. 
In the present simulations, $R_M$ exceeds unity at $p_\mathrm{gas}<2$ bar and peaks near $R_M \approx 70$ at $\sim 0.3$ bar, suggesting that MHD effects could be dynamically important in this region.
An approximate representation of these effects could be added in the model by modifying the magnitude of the magnetic field 
by a first-order correction, e.g., $B=B_\mathrm{dip}\sqrt{1+(\epsilon R_M)^2}$, where $\epsilon$ ($0<\epsilon\leq1$) is a parametrization which captures the enhancement of magnetic coupling when atmospheric induction becomes important.
Here the balance between induction and diffusion was applied to estimate the induced magnetic field $B_\mathrm{ind}\sim \epsilon R_M B_\mathrm{dip}$.
$\epsilon$ can be calibrated by comparing the magnetic field profiles to MHD results (e.g., \citealt{RogersKomacek2014,Soriano-Guerrero2025}). 
\paragraph{Dipole magnetic field:}
The assumption of a dipolar planetary field remains reasonable as a first-order approximation for WASP-18 b. 
The morphology of the magnetic field depends on the balance between Coriolis and inertial forces in the convecting interior, characterized by the Rossby number $Ro=\frac{U}{\Omega L}$ \citep{ChristensenAubert2006}. 
Taking characteristic values of $U\sim0.1$ m\,s$^{-1}$ for the convective velocity, $L\sim5\times10^7$ m for the Jupiter-like shell thickness \citep{Elias-Lopez2025}, and $\Omega=7.7\times10^{-5}$ rad\,s$^{-1}$ for the angular velocity of WASP-18 b, yields $Ro\sim3\times10^{-5}$. 
This value lies below the critical threshold of $Ro\sim0.1$, consistent with a dipole-dominated dynamo regime. 
Higher Rossby numbers ($Ro>0.1$)  indicate weaker, multipolar fields.  
However, $U$ and $L$ are uncertain and model-dependent, and the dynamo region of WASP-18 b is not directly constrained. 
Internal heating due to tides or Ohmic dissipation may affect the thickness and structure of the dynamo shell, which might potentially favor a multipolar field (see, e.g., \citealt{stanleybloxham2006}). 
While the dipole assumption is reasonable for modeling atmospheric drag, the possibility of a multipolar magnetic field cannot be ruled out.
\section{Summary}\label{sec:summary}
In this study, a physically motivated model for the anisotropic magnetic drag is presented to investigate how  an anti-aligned planetary dipole magnetic field influences the atmospheric circulation of (U)HJs. 
This approach addresses the main research questions of (1) how magnetic coupling and the resulting drag modifies the wind circulation and hotspot shift, (2) to what extent the  magnetic drag disrupts the equatorial superrotation and (3) how it alters heat distribution between the day- and nightside and the climate characteristics.  
By connecting the anisotropic drag approach to ionospheric conductivities and magnetization, the approach provides a theoretical framework for linking MHD processes to climate characteristics in UHJ atmospheres.
Therefore, the magnetic drag parametrization is derived from a multi-fluid description in which electrons and ions are treated as magnetized species in steady-state momentum balance, collisionally coupled to the neutrals. The resulting ion and electron drift velocities relative to the neutrals are used to determine anisotropic drag in terms of Hall and Pedersen components acting on the neutral flow.
The parametrization is implemented in the meridional and zonal momentum equations as well as in the temperature evolution equation of the cloud-free 3D GCM \texttt{ExoRad}, and applied to WASP-18 b. 
The results are compared with previously published drag treatments ('uniform drag' and 'active drag') and with simulations with an absent magnetic field influence.
They demonstrate that the atmospheric circulation in particular the presence and width of the equatorial superrotating jet strongly depend on the magnetic drag. 
The main findings are:
\begin{itemize}
    \item Anisotropic drag primarily dissipates kinetic energy and slows the flow on the dayside, while still allowing an efficient eastward equatorial jet to be maintained on the nightside in contrast to uniform drag that suppresses overall wind flow. The presence or absence of anisotropic drag does not affect the width of the nightside jet but it affects the horizontal velocity of the jet.
    \item Active drag completeley decouples the dayside and nightside circulation in the upper atmosphere, leading to a slower nightside jet, a polar flow at the dayside, and a weaker eastward shift of the day-night boundary compared to the case when the magnetic field is absent.
   \item Anisotropic drag  changes  the dayside flow in particular for $p_\mathrm{gas}\leq10^{-3}$~bar. The inclusion of the Hall term and the meridional drag component (Pedersen) leads  to a deflection of the flow perpendicular the dipole field. The result is a direct flow (albeit at weaker level compared to uniform drag) from the dayside towards the nightside at the evening terminator at all latitudes and at the morning terminator at high latitudes ($>|30^{\circ}|$). 
   \item  Identifying the wind flow direction over the terminators is thus important to identify the coupling regime (uniform: diverging flow from the substellar point, active: strong decoupling at terminator regions, anisotropic: only decoupled at equatorial morning terminator, otherwise dayside-to-nightside flow.
   \item Because anisotropic drag includes more drag effects, this leads to an overall stronger damping of wind flow with smaller maximum velocities compared to the active drag. 
    \item With anisotropic drag the evening–morning terminator averaged temperature difference at $p_\mathrm{gas}=10^{-1}$ bar is stronger compared to the case without magnetic drag with the evening terminator being 20\% hotter than the morning terminator.
    \item Both active and anisotropic drag reshape the circulation such that two hotspots appear north and south of the equator, with smaller eastward shifts compared to the case when magnetic drag is absent. The results show that local variations in drag forces and energy can create observable asymmetries, producing both primary and secondary hotspot regions.
\item The morning-to-evening terminator gas temperature difference, $\Delta T_{\rm gas}$, at $p_\mathrm{gas}=10^{-3}$~bar (Tab.~\ref{tab:drag_summary}), is crucial to identify day-to-night side flow patterns arising with different drag treatments: Uniform drag with radial flow exhibits almost no contrast ($\Delta T_{\rm gas} \approx 0$~K), active drag with polar flow exhibits high contrast ($\Delta T_{\rm gas}=145$~K) and anisotropic drag due to partial horizontal flow towards the nightside exhibits lower contrast ($\Delta T_{\rm gas}\approx100$~K).
\end{itemize}
While the parametrization of the magnetic drag includes the influence of Pedersen and Hall drag, it neglects several induction processes such as dynamo action, ambipolar diffusion, and ion/electron pressure effects. These processes can become likely important in the partially ionized upper atmosphere, where $R_M \gtrsim 1$, and may strongly determine both the efficiency and anisotropy of magnetic drag. Observationally, such uncertainties might impact predictions of phase curve amplitudes, hotspot offsets, transit asymmetries, and Doppler wind signatures.
Because the model does not include the self-consistent evolution of the magnetic field, and thus neglects inductive feedback, it likely underestimates magnetic stresses. It is therefore not valid in atmospheric regions where $R_M \gg 1$. However, for $R_M \lesssim 1$, the parametrization provides reasonable estimates of the magnitude and direction of magnetic drag.
Although, the approach neglects the polarization electric field which arise to ensure global current closure, it remains a suitable and computationally efficient method for calculating the first-order effects of magnetic fields in a GCM.
Understanding these effects and to what extent they affect observations can  provide a constrain of the magnetic field strengths of specific UHJs in future work.
\bibliographystyle{aa} 
\bibliography{bib_wasp_18b} 
\begin{appendix}
\nolinenumbers
\section{Active magnetic drag treatment}\label{ap:active_drag}
Here, the active magnetic drag treatment is reviewed, in which the interaction between atmospheric winds and magnetic fields is modeled by calculating the induced currents and resulting Lorentz forces.
This approach was suggested by \cite{Perna2010} and first implemented in hot Jupiter simulations by \cite{RauscherMenou2013}. 
It was later extended to UHJs by \cite{Beltz2022}. 
In contrast to the uniform drag treatment, the active drag approach incorporates spatially and temporally varying diffusivity, providing a more physically realistic representation of magnetic drag in atmospheres that are only partially ionized.
The model is based on the following  assumptions: (a) Planetary magnetic field is a dipole aligned with the planet's rotation axis, (b) Zonal  winds dominate the atmospheric circulation, (c) Magnetic Reynolds number is small ($R_M\ll1 $), so the field is not significantly distorted by magnetic advection and is diffusion-dominated, (d) Since the system is diffusion-dominated other effects such as Hall and ambipolar diffusion are neglected.
Under these assumptions, the non-ideal MHD induction equation simplifies to:
\begin{equation} \label{eq:ind_general}
\partial_{t}\underline{B}=\nabla\times\left(\underline{v}\times\underline{B}-\eta \nabla\times\underline{B}\right),
\end{equation}
where $\underline{B}$~[T] is the magnetic field, $\underline{v}$~[$\frac{m}{s}$] the plasma bulk velocity,  and $\eta(r,\theta,\phi)=\frac{1}{\mu_0\sigma(r,\theta,\phi)}$ [$\frac{m^2}{s}$] is the magnetic diffusivity with the magnetic permeability $\mu_0$[$\frac{N}{A^2}$] and the electrical conductivity $\sigma$ [$\frac{S}{m}$].
Here, $\sigma$ corresponds to the parallel conductivity $\sigma_\parallel$ in the model of \cite{Perna2010} (see also \citealt{Christie2025}).
For the model atmosphere used here, this agrees well with the Pedersen conductivity $\sigma_P$ on the dayside at higher pressure levels ($p_\mathrm{gas}>10^{-3}$ bar), but $\sigma_P$ starts to deviate from $\sigma_\parallel$ at lower pressures ($p_\mathrm{gas}\leq10^{-3}$ bar). Therefore, using $\sigma=\sigma_P$ is more appropriate
for capturing the resistive effect that occurs when the gyrofrequency exceeds the collision frequency.
\cite{Perna2010} uses the velocity of neutrals rather than the plasma bulk velocity ($\underline{v}=\underline{v}_n$), because the plasma is weakly ionized and collisions between ions/electrons and neutrals are frequent.
The neutral gas dominates the mass and momentum budget which leads to ions/electrons being strongly coupled to the neutrals through collisions.
The diffusive term introduces spatial gradients of $\eta$, which must be carefully treated in planetary atmospheres where conductivity varies both radially and horizontally \citep{Perna2010}.
All vector components and derivatives are expressed within the right-handed spherical coordinate system defined in Sec. \ref{ch:momentum_exchange}. 
Note that this approach differs from standard formulations, in which colatitude is often used.
The diffusive term is given by:
\begin{align} \label{eq:ind_eq}
\nabla\times(\eta \nabla\times\underline{B})=&-\eta\nabla^2\underline{B}+\frac{\partial\eta}{\partial r}\underline{e}_r\times(\nabla\times\underline{B})+\frac{1}{r}\frac{\partial\eta}{\partial \theta}\underline{e}_\theta\times(\nabla\times\underline{B}) \nonumber\\
&+\frac{1}{r \cos\theta}\frac{\partial\eta}{\partial \phi}\underline{e}_\phi\times(\nabla\times\underline{B}).
\end{align}
It is useful to introduce the radial, meridional, and zonal magnetic diffusivity scale heights:
\begin{equation}
    H_{\eta,r}=\frac{\eta}{\partial\eta/\partial r}, \quad H_{\eta,\theta}=\frac{r\eta}{\partial\eta/\partial \theta}, \quad H_{\eta,\phi}=\frac{r \cos\theta\,\eta}{\partial\eta/\partial \phi}.
\end{equation}
These scales can be used to evaluate whether horizontal gradients can be neglected relative to the radial one.
The calculated scale heights from the model atmosphere showed that the horizontal scale heights dominate the radial scale height (i.e., $H_{\eta,\theta},H_{\eta,\phi}>H_{\eta,r}$) and therefore horizontal diffusivity gradients in Eq.~\ref{eq:ind_eq} can be neglected.
In order to estimate the relative importance of the advective (first term on the right-hand side of Eq.~\ref{eq:ind_general}) and diffusive terms (second term on the right-hand side of Eq.~\ref{eq:ind_general}) the magnetic Reynolds number $R_M$ is used for dimensional analysis.
The typical length scale for $R_M$ has to be chosen carefully (see therefore e.g., \citealt{Liu2008} and \citealt{Perna2010}).
The pressure scale height $H_p=\frac{R_dT_\mathrm{gas}}{g}$~[m] is smaller than $H_{\eta,r}$  for most of the upper model atmosphere ($p_\mathrm{gas}<0.1$ bar) and similar in magnitude in the lower atmosphere, therefore  $H_p$ is used to calculate the magnetic Reynolds number:
\begin{equation}\label{eq:reynolds}
    R_M=\frac{vH_{p}}{\eta}.
\end{equation}
Note that $R_M$ may exceed unity in the model atmosphere showing that this approach is not suitable to model such hot atmospheres.
When $R_M\ll 1$, the system is in quasi-steady-state and time derivatives can be neglected ($\partial_t=0$).
Assuming negligible dynamo action, the magnetic field is decomposed into a dipolar background ($\underline{B}_\mathrm{dip}$) and an induced component ($\underline{B}_\mathrm{ind}$)(see, e.g., Batygin and Stevenson 2010):
\begin{equation}
    \underline{B}=\underline{B}_\mathrm{dip}+\underline{B}_\mathrm{ind}, \quad \nabla\times \underline{B}_\mathrm{dip}=0.
\end{equation} 
Since the induced field is typically small when $R_M\ll1$, the induction equation is linearized around the background field:
\begin{equation}    \nabla\times(\underline{v}\times\underline{B}_\mathrm{dip})=\nabla\times(\eta \nabla\times\underline{B}_\mathrm{ind}).
\end{equation}
Using Amp\`{e}re's law ($\underline{j}_\mathrm{ind}=\frac{1}{\mu_0}\nabla\times\underline{B}_\mathrm{ind}$), the induction equation becomes:
\begin{equation}   
\nabla\times(\eta \mu_0\underline{j}_\mathrm{ind})=\nabla\times(\underline{v}\times\underline{B}_\mathrm{dip})
\end{equation}
Focusing on the zonal ($\phi$) component and assuming dominant horizontal currents (i.e., $j_r\ll j_\theta$), the dominant balance becomes:
\begin{eqnarray} 
\mu_0\partial_r(r\eta j_{\mathrm{ind},\theta})=
\partial_r(rv_\phi B_{\mathrm{dip},r})+\partial_\theta(v_\phi B_{\mathrm{dip},\theta}).
\end{eqnarray}
Integrating over radius from the layer of interest to the model top at radius $R$, an expression for the meridional current is obtained:
\begin{align}
j_{\mathrm{ind},\theta}(r,\theta,\phi)=&\frac{R\eta(R,\theta,\phi)}{r\eta(r,\theta,\phi)} j_{\mathrm{ind},\theta}(R,\theta,\phi)\nonumber\\
&-\frac{1}{\mu_0r\eta(r,\theta,\phi)}\int_{r}^{R} dr' \left(\partial_r'(r'v_\phi B_{\mathrm{dip},r})+\partial_\theta(v_\phi B_{\mathrm{dip},\theta})\right).
\end{align}
The first term on the right-hand side is a boundary current in the uppermost modeled level and is set to zero for simplicity (Perna et ail 2010).
Furthermore, it is assumed that the radial length scale of diffusivity variations ($H_{\eta}(r)$), is smaller than the characteristic meridional length scales of the zonal wind and magnetic field. 
As a result, radial gradients dominate over meridional ones, allowing to neglect $\frac{1}{r}\partial/\partial\theta$ terms in comparison to $\partial/\partial r$ in the equation above.
Assume that $\partial_r'(r'v_\phi B_{\mathrm{dip},r})$ slowly varies with radius and that the atmosphere is very thin compared to the radius of the planet, then it can be approximated by
\begin{eqnarray}
j_{\mathrm{ind},\theta}(r,\theta,\phi)&=&-\frac{1}{\mu_0r\eta(r,\theta,\phi)}\int_{r}^{R} dr' \left(\partial_r'(r'v_\phi B_{\mathrm{dip},r})\right) \nonumber\\
&\approx& -\frac{(R-r)}{\mu_0r\eta(r,\theta,\phi)} \partial_r(rv_\phi B_{\mathrm{dip},r})\nonumber\\ &\approx& -\frac{v_\phi}{\mu_0\eta(r,\theta,\phi)} B_{\mathrm{dip},r}.
\end{eqnarray}
The resulting Lorentz force appears in the zonal momentum equation (where for the  mass density $\rho$ the neutral gas mass density $\rho_n$ was assumed) as:
\begin{equation}
    \rho\frac{d}{dt}v_\phi\approx-j_{\mathrm{ind},\theta} B_{\mathrm{dip},r}
\end{equation}
Substituting the idealized expression for the meridional current
$j_{\mathrm{ind},\theta}=- v_\phi B_{\mathrm{dip},r}/ (\mu_0\eta)$ gives a drag force proportional to the zonal wind velocity:
\begin{equation}
    \rho\frac{d}{dt}v_\phi\approx\frac{1}{\eta\mu_0 }v_\phi B_{\mathrm{dip},r}^2.
\end{equation}
Assuming a dipolar field with radial component $B_{\mathrm{dip},r}=2B_0\sin\theta\frac{R_p^3}{r^3}$ the equation becomes:
\begin{equation}
    \rho\frac{d}{dt}v_\phi\approx\frac{1}{\eta \mu_0}v_\phi \left(2B_0\sin\theta\frac{R_p^3}{r^3}\right)^2.
\end{equation}
Furthermore, the magnetic field strength is assumed to be independent of radius, simplifying the radial variation of $\tau_\mathrm{drag}$.
This yields a magnetic drag timescale of the form:
\begin{equation}\label{eq:dragt}
\tau_\mathrm{drag}=\frac{\mu_0\rho\eta}{B_{\mathrm{dip},r}^2}.
\end{equation}
$\tau_\mathrm{drag}$ \footnote{Due to the quadratic dependence on the radial magnetic field component  ($B_{\mathrm{dip},r}^2$), the drag timescale $\tau_\mathrm{drag}$ is insensitive to whether the dipole magnetic field is aligned or anti-aligned with the rotation axis.} is used in this study for the 'active drag' treatment.
Note that this approach differs from the parametrized model used in \cite{RauscherMenou2013} and \cite{Beltz2022}, where the drag force is based on local flow geometry, magnetic field strength, and diffusivity via
\begin{equation} \label{eq:dragtPe}
\tau_\mathrm{drag,Perna}=\frac{\mu_0\rho\eta}{B_0^2|\sin\theta|}.
\end{equation}
The latitudinal dependence of the radial magnetic flux density is not fully maintained in their formulation of the drag timescale and interprets drag as being proportional to the component of the wind perpendicular to the magnetic field lines.
At the equator ($\theta=0$°), the magnetic drag vanishes ($\tau_\mathrm{drag}=\tau_\mathrm{drag,Perna}\rightarrow\infty$).

The energy equation is also modified by the kinetic energy which is lost due to the magnetic drag and  returned to the atmosphere in form of localized ohmic heating with 
\begin{equation}\label{eq:qfricactive}
    q_\mathrm{fric,active}=v_\phi^2/\tau_\mathrm{drag}.
\end{equation}
Note that in the 'uniform drag' treatment the heating rate per unit mass is given by:
\begin{equation}\label{eq:qfricuni}
    q_\mathrm{fric,uniform}=(v_\phi^2+v_\theta^2)/\tau_\mathrm{drag,uniform}.
\end{equation} 
While the active drag approach is more physically motivated than the empirical Rayleigh drag, it is still an  approximation. 
In particular, the assumption of dominant zonal winds, weak induced fields, and $R_M\ll1$ may not hold in all regions of the atmosphere as shown for WASP-18 b.
However,  this model was still used for simulations of (U)HJs where the condition of $R_M\ll1$ was not given (e.g., \citealt{Beltz2022,Coulombe2023}).
\subsection{Thermal ionization}
Due to extreme temperatures on the dayside ($>3000$ K), thermal collisions are sufficiently energetic to ionize atoms and molecules and form an ionosphere. 
Thermal ionization rather than photoionization is the dominant process creating strong ionospheric effects in our model, influencing the atmospheric dynamics and magnetic interactions.
However, \cite{Koskinen2014} showed that photoionization can dominate over thermal ionization on the dayside atmosphere above the 100 mbar level.
In order to determine the thermal ionization fraction $x_{e,\mathrm{Saha}}$ in a gas in local thermodynamic equilibrium, models of \cite{Menou2012},  \cite{RauscherMenou2013} and others adopted the approximate solution of the Saha equation from \cite{Sato1991}.
Furthermore, they assumed that the ionization fraction is small ($x_{e,\mathrm{Saha}}\ll 1$). 
Therefore, only single ionization from the ground state is considered \citep{Menou2012}.
The Saha equation for a plasma in thermodynamic equilibrium is given by
\begin{equation}\label{eq:Saha}
    \frac{n_{i,j}^+n_e}{n_{n,j}}=\left(\frac{m_ek_BT}{2\pi \hbar^2}\right)^\frac{3}{2}\frac{2Z_{i,j}}{Z_{n,j}}\exp\left(\frac{-I_{i,j}}{k_BT}\right),
\end{equation}
with the electron mass $m_e$~[kg], the Boltzmann's constant $k_B$~[$\frac{J}{K}$], the gas temperature $T_\mathrm{gas}$~[K], the reduced Plank's constant $\hbar$[Js], the electron number density $n_e$~[m$^{-3}$], the ionized number density of the $j$th chemical species $n_{i,j}^+$~[m$^{-3}$], the atom number density $n_{n,j}$~[m$^{-3}$] of the $j$th species, their partition functions $Z_{i,j}$ and $Z_{n,j}$, and  the first ionization potential $I_{i,j}$~[J] of the $j$th species.
For the calculation of $x_e$ the first 28 elements from the periodic table (hydrogen to nickel) are taken into account \citep{Menou2012, RauscherMenou2013}.
The first ionization potential $I_{i,j}$ ranges from 7$\times$10$^{-19}$~J (for K) to 4$\times$10$^{-18}$~J (for He)
The partition functions $Z_{i,j}$ and $Z_{n,j}$ depend on the temperature in a complex way \citep{Sato1991}. 
For simplicity, $Z_{i,j}$ and $Z_{n,j}$ are set to unity according to \cite{Menou2012} since the specific atmospheric composition is unknown.
\cite{Menou2012} and \cite{RauscherMenou2013} use a sum of approximate Saha expressions across multiple elements.
They evaluate ionization fractions for each element ($x_{e,\mathrm{Saha},j}=n_{i,j}^+/n_\mathrm{tot,j}$) independently using uncoupled Saha equations.
The total number density of the $j$th element is given by $n_\mathrm{tot,j}=n_{n,j}+n_{i,j}^+$.
They assume that only the $j$th species contributes to the electrons and therefore use $n_e=n_{i,j}^+$, which is not mathematically consistent, but here a useful simplification  in a weakly ionized plasma in order to derive $x_{e,\mathrm{Saha}}$. 
The Saha equation \ref{eq:Saha} can be described in terms of the ionization fraction for each ion species: 
\begin{equation}\label{eq:Saha_final}
    \frac{x_{e,\mathrm{Saha},j}^2}{1-x_{e,\mathrm{Saha},j}}=\frac{2}{n_{n,j}}\left(\frac{m_ek_BT_\mathrm{gas}}{2\pi \hbar^2}\right)^\frac{3}{2}\exp(-I_j/k_BT_\mathrm{gas}),
\end{equation}
where 
$n_{n,j}=\frac{a_j}{a_H}n_n$ is the number density of the element $j$, which is calculated by assuming solar abundance ($a_j$) for element $j$ and the hydrogen abundance ($a_H$), $n_n$ is the total number density of the neutrals. 
The numbers for the solar abundances were taken from \cite{Asplund2021}.
Then the total ionization fraction can be described as
\begin{equation}
    x_{e,\mathrm{Saha}}=\frac{n_e}{n_\mathrm{tot}}\approx\sum_{j=1}^{28}\frac{n_{i,j}^+}{n_n}\approx\sum_{j=1}^{28}\frac{n_{n,j}}{n_n}x_{e,\mathrm{Saha},j},
\end{equation}
with the total number density of atoms/particles in all ionization states (not including electrons as independent atoms):
\begin{eqnarray}
    n_\mathrm{tot}=\sum_{j=1}^{28}(n_{n,j}+n_{i,j}).
\end{eqnarray}
It was assumed here that the plasma is weakly ionized ($n_{i,j}^+\ll n_{n,j}$).
Note that Eq.~\ref{eq:Saha_final} is slightly different from the derived equations in e.g, \cite{RauscherMenou2013}.

A more self-consistent treatment of the ionization balance is adopted for the anisotropic drag formulation using the GGchem model, compared to the approximation used by \cite{RauscherMenou2013} (see Sec. \ref{ch:momentum_exchange}).
The difference between the two approximations (simplified Saha and \text{GGchem} approximation) is small for $p_\mathrm{gas}<0.03$ bar, deeper in the atmosphere  it starts to deviate with the largest deviation around 6 bar by a factor larger than 10 for the \texttt{GGchem} approach.  
\subsection{Magnetic diffusivity}
The calculation of the ionization fraction $x_e$ allows to evaluate  the magnetic diffusivity $\eta$ in the atmosphere via \citep{Perna2010,Menou2012}
\begin{equation}\label{eq:eta}
    \eta=\frac{1}{\mu_0\sigma_\parallel}\approx 0.023\frac{\sqrt{T_\mathrm{gas}}}{x_e}\, [\mathrm{m^2s^{-1}}].
\end{equation}
This expression was derived from the approximated momentum transfer due to collisions between the electrons and the neutrals with the collision rate \citep{Draine1983}
\begin{equation}
    <\sigma v>_e\approx10^{-19}\sqrt{\frac{128 k_B T_\mathrm{gas}}{9\pi m_e}} \, \mathrm{[m^3 s^{-1}]}.
\end{equation}
$\sigma_\parallel$ was determined by 
\begin{equation}\label{eq:el_conductivity}
    \sigma_\parallel=\frac{n_e e^2}{m_e n_n <\sigma v>_e}\,\mathrm{[S\, m^{-1}]}.
\end{equation}
\section{Alternative derivation of the anisotropic drag}\label{ap:christe}
The model of \cite{Christie2025} extends the active drag model of \cite{Perna2010} and derives an approximation of the Lorentz force   which includes both the Pedersen and Hall effect through anisotropic conductivities.
The drag force is expressed in terms of the anisotropic ionospheric conductivities for ions and electrons:
\begin{equation}\label{eq:drag_conductivities}
    \underline{F}_\mathrm{drag}=\underline{j}\times\underline{B}_\mathrm{dip}=-B_\mathrm{dip}^2\sigma_P\underline{v}_{n,\perp}+B_\mathrm{dip}^2\sigma_H(\underline{v}_n\times\underline{\hat{b}})
\end{equation}
with the Pedersen ($\sigma_P$) and Hall conductivities ($\sigma_H$) for plasma  consisting of one ion and electron species: 
\begin{eqnarray} \label{eq:hall_pedersen}
    \sigma_P&=& \frac{n_ie^2}{m_i\nu_{in}(1+k_i^2)}+\frac{n_ee^2}{m_e\nu_{en}(1+k_e^2)}=\sigma_{P,i}+\sigma_{P,e} \\
    \sigma_H&=&\frac{n_ee^2k_e}{m_e\nu_{en}(1+k_e^2)}-\frac{n_ie^2k_i}{m_i\nu_{in}(1+k_i^2)}= \sigma_{H,e}-\sigma_{H,i}.
    \label{eq:hall_pedersen_2}
\end{eqnarray}
For the parameter regime used for  WASP-18 b simulations, the ionospheric conductivities can be approximated by $\sigma_H\approx \sigma_{H,e}$ and $\sigma_P\approx \sigma_{P,e}$.
For the parameters used here, the comparison between the factors in Eqs.~\ref{eq:drag_conductivities} and \ref{eq:drag_final} shows that they are the same in magnitude, i.e., $\sigma_{P}\approx K_P/B_\mathrm{dip}^2$ and $\sigma_{H}\approx K_H/B_\mathrm{dip}^2$.
The main difference between this approach and the parametrized anisotropic drag  presented in Sec. \ref{sec:parametrized_drag} is the physical origin of the drag term.
In Sec. \ref{sec:parametrized_drag}, the magnetic drag exerted on the neutrals is directly calculated from the collisional momentum transfer between the plasma and the neutral fluid, whereas the method of \cite{Christie2025} describes the Lorentz force acting on the plasma fluid, which is then balanced by the drag force exerted by the neutrals.
The method to  parametrize the collisional drag force on the neutrals via the differential velocities ($\Delta \underline{v}_s$) is more direct,  since the Lorentz force does not appear in the neutral momentum equation (Eq.~\ref{eq:neutrals}).
Neutrals experience magnetic effects only though collisions with the charged species.
Furthermore, \citet{Christie2025} include both horizontal and vertical components of the magnetic drag force in their non-hydrostatic model, whereas in this work the vertical drag component is neglected since a hydrostatic framework in the vertical direction is applied. 
However, note that in Eq.~\ref{eq:drag_conductivities} the conventional conductivities are defined in the plasma's rest frame (Eqs.~\ref{eq:hall_pedersen}, \ref{eq:hall_pedersen_2}).
In order to use it in the neutral gas's rest frame, the conductivities have to be transformed into effective conductivities which are more complex and are given in section 4 of \cite{Song2001}.
However, the effective conductivities become the same as  the conventional conductivities (here $\sigma_H\approx \sigma_{H,e}$ and $\sigma_P\approx \sigma_{P,e}$) under the following approximations \citep{Song2001}: $\nu_{in}\nu_{en}\gg\omega_i\omega_e$, $\nu_{en}\gg \omega_e$, and negligible  collisions between electrons and ions. 
For the parameter regime on the dayside of the simulations WASP-18 b these conditions are fulfilled and the drag formulation in Eq.~\ref{eq:drag_conductivities} is a suitable approach for the magnetic drag to be applied in \texttt{ExoRad}.
The energy equation is accordingly modified with the heating rate:
\begin{equation}
    q_\mathrm{fric}=\frac{1}{\rho_n}\sigma_PB_\mathrm{dip}^2\underline{v}_{n,\perp}^2.
\end{equation}
\section{Comparison of different ion-neutral drift velocities}\label{ap:savel}
\begin{figure*}
\noindent\includegraphics[width=1\textwidth]{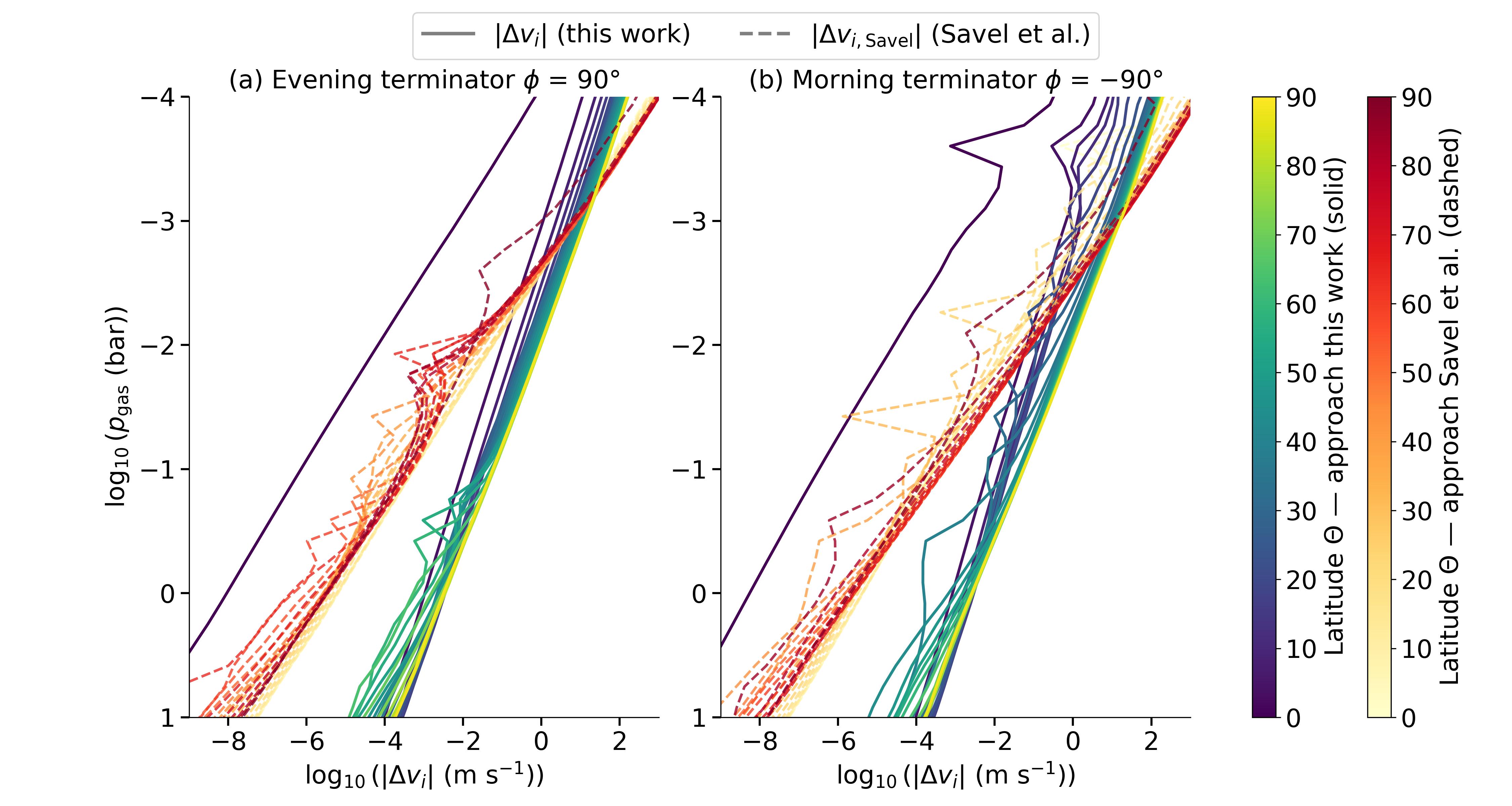}
\caption{Ion–neutral drift velocities as a function of pressure and latitude at the evening terminator (a): $\phi=90$°) and morning terminator (b): $\phi=-90$°) for two different approaches.
Solid lines show $|\Delta \underline{v}_i|$ from the present formulation (Eq.~\ref{ref:vel_s}) and dashed lines follow \cite{Savel2024} (Eq.~\ref{eq:savel}). Colors denote the latitude. The drift velocities are approximated using neutral wind fields from a simulation without magnetic drag and a dipolar magnetic field (Eq.~\ref{eq:dipole}) with $B_0=5$~G in both approaches. The results demonstrate the latitudinal dependence introduced by magnetic field geometry and the effect of anisotropic drag on the ion-neutral drift velocity.} 
\label{Fig:savel_profile}
\end{figure*}
As already pointed out by \cite{Christie2025} the relative velocity between ions and neutrals in the atmosphere derived using the anisotropic drag approach  differs  from that applied in the toy model of  \cite{Savel2024}.
While the drift velocity between ions (index $s=i$) and neutrals is given in Eq.~\ref{ref:vel_s}, \cite{Savel2024} assumes that the conductivity tensor can be expressed as a scalar in this system and therefore applies the approach from \cite{Perna2010} for the meridional current density $j_{\mathrm{ind},\theta}\sim v_\phi B_{\mathrm{dip}}/ (\mu_0\eta)$  resulting in a different drift velocity:
\begin{equation}\label{eq:savel}
    |\Delta \underline{v}_{i,\mathrm{Savel}}|=\left|\frac{\underline{j}\times\underline{B}}{n_im_i\nu_{in}}\right|\approx \frac{|v_\phi| B_{\mathrm{dip}}^2}{n_im_i\nu_{in}\mu_0\eta}.
\end{equation}
In this approach, the assumed electrical conductivity is the parallel conductivity (Eq. \ref{eq:el_conductivity}). This value begins to deviate from the Pedersen conductivity above the M1 region (see Fig.~4 in \citealt{Koskinen2014} or Fig.~2 in \citealt{Christie2025}) leading to larger drift velocities in the Savel et al. approach. The M1 region is defined as the area where particles are not coupled to the magnetic field lines and both ion and electron magnetizations (Eq. \ref{eq:magnetization}) satisfy $k_i, k_e\ll1$.
A discussion on the differences of the two different approaches is given in \cite{Christie2025}.
Fig.~\ref{Fig:savel_profile} compares the approximated ion–neutral drift velocities derived in this work with those predicted by the approach of \cite{Savel2024}. 
Both approaches show a similar monotonic increase in drift speed with decreasing pressure, indicating that collisional coupling becomes weaker as the density decreases. 
However, the Savel et al. approach predicts larger drift velocities by up to several orders of magnitude for $p_\mathrm{gas}\leq10^{-4}$~bar, the pressures which are generally probed in high-resolution transmission spectroscopy \citep{Savel2024}. 
The simulation domain extends up to pressures of $p_\mathrm{gas}\sim10^{-4}$~bar. 
Extrapolating the derived ion–neutral drift to lower pressures suggests that relative velocities could reach values of order km~s$^{-1}$, and may therefore be detectable with high-resolution spectroscopy in the photosphere. 
The equatorial ($\theta=0$°) $|\Delta \underline{v}_i|$ profile behaves differently from higher latitudes due to the dipolar magnetic field geometry,  where the radial magnetic field vanishes at the equator and the ion–neutral drift velocity is purely zonal. 
The differences in the profiles and approaches highlight the magnetic field geometry dependence. 
Differences also occur between the morning and evening terminators due to the feedback of the anisotropic drag on the atmospheric flow velocity.
These dependencies emphasize the need for accurate physical models to interpret ion–neutral drift measurements and to constrain exoplanetary magnetic field strengths.
\end{appendix}

\begin{acknowledgements}
      A.B. thanks Maxim Khodachenko for useful discussions.
      The post-processing of GCM  data has been performed with gcm-toolkit \citep{Schneider2022}. The simulations (project id 72245) were performed on the Austrian Scientific Computing (ASC) infrastructure, in particular the Vienna Science Cluster (VSC). We thank the referee for very useful comments that improved this manuscript.
\end{acknowledgements}

\end{document}